\documentclass[10pt]{article}
\usepackage{amsmath}
\usepackage{graphicx}
\usepackage{color}
\usepackage{orcidlink}
\usepackage{hyperref}
\hypersetup{colorlinks=true, linkcolor=blue, citecolor=blue, urlcolor=blue}
\usepackage{caption}
\usepackage{subcaption}
\usepackage{setspace}
\usepackage{multirow}
\usepackage{multicol}
\usepackage{amssymb}
\RequirePackage[numbers,sort&compress]{natbib}
\begin{document}
\baselineskip=14pt

\begin{center}
\LARGE{Quantum-Corrected Deformation of RN-AdS Black Holes with a Cloud of Strings Immersed in Quintessential-like Fluid }
\par\end{center}

\vspace{0.3cm}

\begin{center}
{\bf Faizuddin Ahmed\orcidlink{0000-0003-2196-9622}}\footnote{\bf faizuddinahmed15@gmail.com}\\
{\it Department of Physics, Royal Global University, Guwahati, 781035, Assam, India}\\
\vspace{0.2cm}
{\bf Abdelmalek Bouzenada\orcidlink{0000-0002-3363-980X}}\footnote{\bf abdelmalekbouzenada@gmail.com (Corresp. author)}\\ 
{\it Laboratory of Theoretical and Applied Physics, Echahid Cheikh Larbi Tebessi University 12001, Algeria}\\

\vspace{0.4cm}

\end{center}

\begin{abstract}
Inspired by Loop Quantum Gravity (LQG), we investigate the Reissner–Nordström (RN) black hole (BH) solution coupled with a cloud of strings in an anti-de Sitter (AdS) background, surrounded by a quintessence-like fluid. We begin by analyzing the optical properties of the BH through null geodesic motion, deriving an effective potential that governs photon dynamics. This effective potential is central to understanding key phenomena such as photon trajectories, circular null orbits, photon sphere, and the resulting BH shadow. Subsequently, we study the dynamics of neutral test particles by deriving effective potential that describes their motion. Using the potential, we compute the specific energy and specific angular momentum of neutral particles in circular orbits around the BH and analyze the outcomes. We also investigate the innermost stable circular orbits (ISCO) and demonstrate how various geometrical and physical properties influence the radius of ISCO. Furthermore, we explore the thermodynamic properties of the BH solution by deriving key quantities such as the Hawking temperature, entropy, Gibbs free energy, internal energy, and specific heat capacity. Throughout the study, we demonstrate that the geodesic structure, scalar field behavior, and thermodynamic properties are significantly influenced by parameters such as the string cloud, quantum correction, electric charge, the surrounding quintessence-like fluid, and the AdS curvature radius.
\end{abstract}

\section{Introduction}\label{S1}

The study of black hole (BH) solutions remains central to modern theoretical and observational physics, offering insights into the structure of space-time, gravitational dynamics, and the intersection of classical and quantum gravity theories \cite{k1,k2,k3,k4,k5,k6}. These investigations are particularly relevant in extreme regimes where general relativity (GR) may require modifications to account for phenomena beyond its classical framework. Key considerations include the preservation of Lorentz symmetry at low energies and the emergence of quantum gravitational effects near the Planck scale \cite{k7}. In such high-energy domains, classical space-time descriptions break down, necessitating alternative approaches such as Loop Quantum Gravity (LQG), string theory, and effective field theories. Moreover, the presence of a cosmological constant significantly affects the thermodynamic and optical properties of BHs-impacting temperature, entropy, phase transitions, and gravitational lensing \cite{k10}. BH solutions thus serve as powerful theoretical laboratories for testing gravity and exploring new physics that bridges GR with quantum mechanics.

In the context of the AdS/CFT correspondence \cite{AdS1, AdS2}, which relates gravitational theories in Anti-de Sitter (AdS) space to conformal field theories (CFT) on the boundary, the study of BHs in AdS backgrounds has become foundational in theoretical physics \cite{AdS3}. This duality enables the exploration of BH thermodynamics and quantum features via strongly coupled field theories, particularly relevant in high-temperature regimes such as holographic superconductors \cite{AdS4}. Given that BHs possess well-defined thermodynamic quantities like temperature and entropy \cite{AdS5, AdS6, AdS7}, AdS space-times provide a rich setting for analyzing phase transitions and critical phenomena. Notably, the Hawking-Page transition, describing a thermal shift between AdS space and Schwarzschild-AdS BHs, maps to a confinement-deconfinement transition in the dual CFT \cite{AdS8}. Chamblin {\it et al.} extended this by analyzing charged AdS BHs, revealing van der Waals-like phase behavior akin to liquid–gas systems \cite{AdS9}. These insights were expanded through extended phase space thermodynamics, where the negative cosmological constant is treated as pressure and its conjugate as thermodynamic volume, uncovering phenomena such as reentrant phase transitions, triple points, and criticality \cite{AdS10, AdS11}. This led to the development of BH chemistry, focusing on $P$–$V$ criticality, thermodynamic volume, and quantum corrections in AdS BHs \cite{AdS39, AdS40}. Within this framework, the Reissner–Nordström (RN) AdS BH coupled to a cloud of strings presents an intriguing model, incorporating stringy matter as a topological defect characterized by a deficit angle \cite{AdS12, AdS13}. The string cloud modifies the metric through an additional term proportional to the string parameter, influencing the horizon structure and thermodynamic quantities such as Hawking temperature, entropy, specific heat, and free energy. When embedded in AdS space, such configurations exhibit altered critical behavior and stability conditions \cite{AdS14, AdS15}.

The study of BH thermodynamics deepens our understanding of classical general relativity, quantum mechanics, and statistical physics, offering key insights into the nature of gravity and space-time. BHs are now treated as thermodynamic systems obeying analogues of the four laws of thermodynamics, where mass corresponds to internal energy, surface gravity to temperature, and horizon area to entropy \cite{TH4,TH5,TH6,TH7,TH8,TH9}. Hawking’s discovery of BH radiation confirmed that BHs possess a quantum temperature ($T_H$) and entropy ($S$), linking them to quantum statistical mechanics. In Anti-de Sitter (AdS) space-time, BHs have attracted significant interest due to their role in the AdS/CFT correspondence, which relates a $(\mathcal{D}+1)$-dimensional gravitational theory in AdS space to a $\mathcal{D}$-dimensional conformal field theory (CFT) on its boundary \cite{TH1,TH2,TH3}. This duality allows BH thermodynamic properties to model strongly coupled quantum field systems \cite{TH4,TH5,TH6,TH7,TH8,TH9,TH10}. A key example is the Hawking-Page transition, interpreted in the CFT as a confinement/deconfinement phase change \cite{TH11}. These transitions are marked by discontinuities in response functions like heat capacity \cite{TH12,TH13,TH14}, indicating changes in stability \cite{TH15}. To explore such behavior, geometric thermodynamics introduces Riemannian metrics on the space of equilibrium states. Weinhold’s and Ruppeiner’s metrics, based on second derivatives of internal energy and entropy respectively, help identify phase transitions and microscopic interactions \cite{TH16,TH17,TH18,TH19}, with scalar curvature singularities often aligning with critical points \cite{TH20}. This geometric framework complements traditional thermodynamics, offering deeper insight into BH microstructure and stability in both classical and quantum gravity regimes.

The primary objective of this work is to explore a quantum-corrected deformation of RN-AdS BH solution, coupled with a cloud of strings and immersed in a quintessence-like fluid. This comprehensive investigation encompasses the optical properties, dynamical behavior of neutral test particles, and thermodynamic characteristics of the modified BH space-time. Specifically, we analyze how the geometric and physical parameters-such as the quantum correction term, string cloud, and the quintessence-like fluid affect the photon trajectories, the radius of the photon sphere, the size of the BH shadow, and the stability conditions for circular null geodesics. Additionally, we examine the impact of these parameters on the dynamics of neutral test particles by evaluating the effective radial force and determining the location of the innermost stable circular orbits (ISCO). From a thermodynamic perspective, we derive and analyze key quantities including the Hawking temperature, internal energy, Gibbs free energy, and heat capacity, illustrating how these are influenced by the modifications in the space-time geometry. Through this investigation, we aim to deepen the understanding of how quantum gravitational effects, topological defects (such as a cloud of strings), and exotic energy components (like quintessence) interplay to shape the physical and thermodynamic behavior of BHs. The results present in this study are expected to offer new insights into BH physics, particularly in the context of semi-classical gravity and modified gravity theories, and may provide potential observational implications through gravitational lensing and shadow analysis.

The paper is organized as follows: Section \ref{S1} introduces the theoretical background. Section \ref{S2} presents the quantum-corrected deformation of the RN-AdS black hole with a cloud of strings in a quintessence-like fluid, detailing the modified spacetime geometry. Section \ref{S3} analyzes the black hole’s optical properties, including light behavior and gravitational lensing. Section \ref{S4} examines the dynamics of neutral test particles. Section \ref{S5} investigates thermodynamic properties such as temperature, entropy, and stability. Finally, Section \ref{S6} summarizes the conclusions, physical implications, and future research directions.

\section{Quantum-Corrected RN-AdS BH with CoS immersed in QF }\label{S2}

In this section, we present a static, spherically symmetric quantum-corrected RN-AdS BH solution coupled with a cloud of strings (CoS) and surrounded by quintessence-like fluid (QF). We aim to investigate the optical properties, dynamics of test particles, and thermodynamic properties of this BH solution. Before proceeding with the analysis, we outline the background space-time geometry that forms the basis of the current study.

In Ref. \cite{JMT}, a static and spherically symmetric charged BH solution coupled with a cloud of strings, surrounded by a quintessence field was discussed. The authors investigated thermodynamic behavior and calculate the quasinormal frequencies for a scalar field in this BH space-time. The line element is described by the Eq. (\ref{aa1}) with the metric function given by
\begin{align}
ds^2 &= -A(r) dt^2 + \frac{dr^2}{A(r)} + r^2\,d\Omega^2,\label{aa3}\\
A(r) &=1-\alpha-\frac{2\,M}{r}+\frac{Q^2}{r^2}-\frac{\mathrm{N}}{r^{3\,\omega+1}}-\frac{\Lambda}{3}\,r^2,\label{aa3a}
\end{align}
where $\alpha$ is the string parameter \cite{PSL} and $ d\Omega^2 = d\theta^2 + \sin^2\theta\, d\phi^2 $ represents the unit sphere solid angle.

This formulation incorporates several key parameters: $ M $ is the BH mass, $ \ell $ is the AdS length scale, $ \mathrm{N} $ quantifies the normalization constant with $w$ is the state parameter of quintessence-like fluid, and $ Q $ is the electric charge. Noted that quintessence-like fluid acting as dark energy source was first proposed by Kiselv \cite{VVK}. It is worth noting that null geodesic motion in RN BH surrounded by quintessence was investigated in Ref. \cite{BM}, while a static and spherically symmetric BH with a cloud of strings (Letelier spacetime) immersed in a quintessential fluid was studied in Ref.. \cite{MMDC}.

In Ref.~\cite{ref1}, the authors explored quantum corrections to an AdS-RN BH solution surrounded by a Kiselev-type fluid, which models quintessence or phantom dark energy, and studied its Joule-Thomson expansion in the extended thermodynamic phase space. In Ref.~\cite{ref2}, using the generalized off-shell Helmholtz free energy framework, the thermodynamic topology of quantum-corrected AdS-RN BH in Kiselev space-time was investigated. Additionally, Ref.~\cite{ref3} examined the Weak Gravity Conjecture (WGC) in the context of quantum-corrected AdS-RN BH surrounded by quintessence and analyzed the results. 

The space-time geometry for a quantum-corrected RN-AdS BH, embedded within a cosmological fluid, is characterized by the following spherically symmetric metric~\cite{ref1,ref2,ref3}:
\begin{align}
ds^2 &= -F(r) dt^2 + \frac{dr^2}{F(r)} + r^2\,d\Omega^2,\label{aa1}\\
F(r) &=-\frac{2\,M}{r} + \frac{\sqrt{r^2 - a^2}}{r} + \frac{r^2}{\ell^2_p} - \frac{\mathrm{N}}{r^{3\,\omega+1}} + \frac{Q^2}{r^2}.\label{aa2}
\end{align}

Inspired by these, we consider a static and spherically symmetric quantum-corrected deformation of RN-AdS BH coupled with a cloud of strings and immersed in quintessential-like fluid. The line element describing this BH solution is given by 
\begin{equation}
ds^2 = -f(r)\,dt^2 + \frac{dr^2}{f(r)} + r^2\,d\Omega^2,\label{aa4}
\end{equation}
where the metric function is given by
\begin{equation}
f(r) =-\alpha-\frac{2\,M}{r} + \frac{\sqrt{r^2 - a^2}}{r}- \frac{\mathrm{N}}{r^{3\,\omega+1}} + \frac{Q^2}{r^2}-\frac{\Lambda}{3}\,r^2.\label{aa5}
\end{equation}
For small $ a/r $, the metric function given in Eq.~(\ref{aa5}) can be approximated as,
\begin{equation}
f(r) \approx 1-\alpha-\frac{2\,M}{r}+\frac{Q^2_\text{eff}}{r^2}-\frac{\mathrm{N}}{r^{3\,\omega+1}}-\frac{\Lambda}{3}\,r^2,\quad\quad Q^2_\text{eff}=Q^2-a^2/2.\label{aa6}
\end{equation}

\begin{figure}[ht!]
    \centering
    \includegraphics[width=0.4\linewidth]{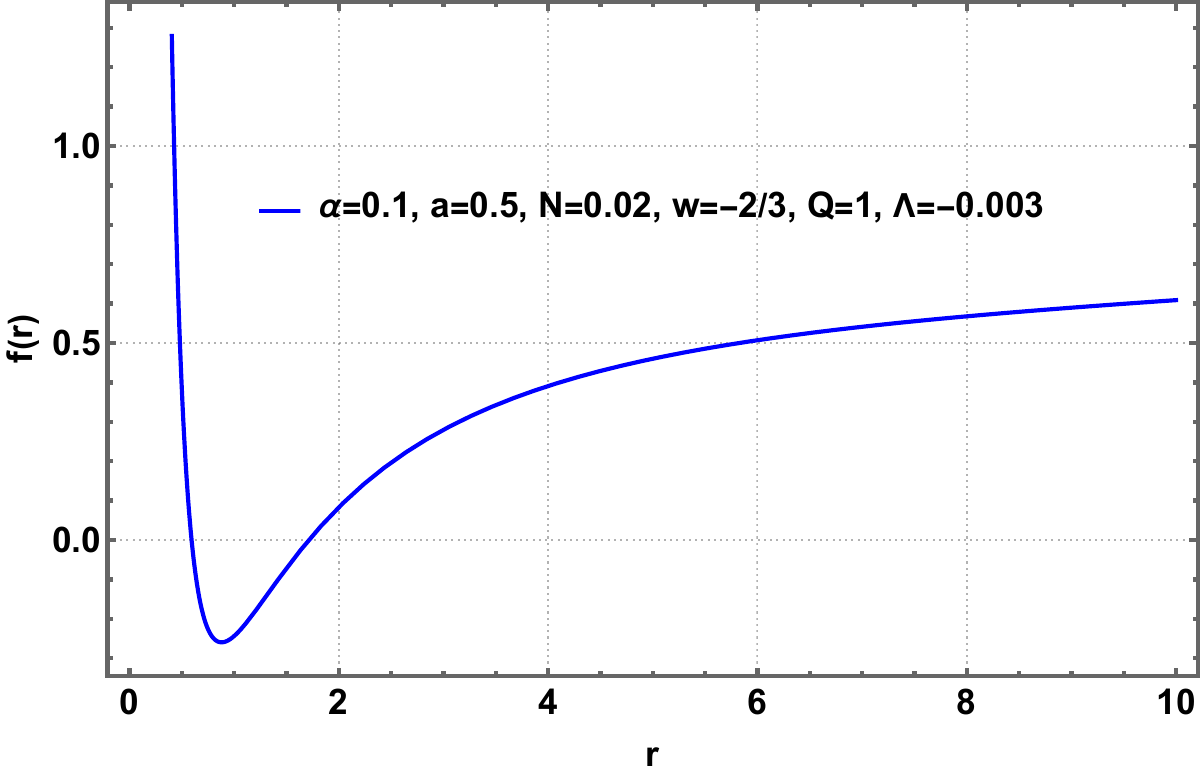}
    \caption{\footnotesize The behavior of the metric function $f(r)$ given in Eq. (\ref{aa6}) as a function of $r$. Here $M = 1$.}
    \label{fig:metric}
\end{figure}

In the limit $ a \to 0 $, which corresponds to the absence of deformation effects in the BH solution, the metric reduces to the well-known RN-AdS space-time coupled with a cloud of strings and surrounded by a quintessence field, as discussed in \cite{ref1, ref2, ref3}. Throughout this manuscript, we adopt the metric function given in Eq.~(\ref{aa6}) to investigate the optical properties, dynamics of neutral particles, scalar perturbations, and thermodynamic behavior of the BH solution under consideration.

\begin{table}[ht!]
\centering
\begin{minipage}{0.45\textwidth}
\centering
\begin{tabular}{|c|c|c|c|c|}
\hline
$a \downarrow \, / \, \alpha \rightarrow$ & 0.05 & 0.10 & 0.15 & 0.20 \\
\hline
0.1 & 2.062 & 2.188 & 2.330 & 2.490 \\
0.2 & 2.090 & 2.216 & 2.357 & 2.517 \\
0.3 & 2.117 & 2.243 & 2.384 & 2.544 \\
0.4 & 2.143 & 2.269 & 2.410 & 2.570 \\
\hline
\end{tabular}
\subcaption*{(a) $Q = 0.5$}
\end{minipage}
\hfill
\begin{minipage}{0.45\textwidth}
\centering
\begin{tabular}{|c|c|c|c|c|}
\hline
$a \downarrow \, / \, \alpha \rightarrow$ & 0.05 & 0.10 & 0.15 & 0.20 \\
\hline
0.1 & 1.448 & 1.608 & 1.776 & 1.958 \\
0.2 & 1.517 & 1.667 & 1.829 & 2.006 \\
0.3 & 1.576 & 1.720 & 1.877 & 2.051 \\
0.4 & 1.629 & 1.769 & 1.922 & 2.094 \\
\hline
\end{tabular}
\subcaption*{(b) $Q = 1.0$}
\end{minipage}
\caption{\footnotesize The BH horizon radius $r_{+}$ for different values of the quantum correction parameter $a$ and the CoS parameter $\alpha$. Here $M=1$, $\Lambda = -0.01$, and $\mathrm{N} = 0.01, w=-2/3$.}
\label{tab:4}
\end{table}

The horizon radius can be determined using the following condition:
\begin{equation}
    f(r)=1-\alpha-\frac{2\,M}{r}+\frac{Q^2_\text{eff}}{r^2}-\frac{\mathrm{N}}{r^{3\,\omega+1}}-\frac{\Lambda}{3}\,r^2=0.\label{aa7}
\end{equation}

In Figure~\ref{fig:metric}, we illustrate the behavior of the metric function $f(r)$ as given in Eq.~(\ref{aa6}). In Table~\ref{tab:4}, we present the numerical values of the BH horizon radius $r_{+}$ for various values of the cloud of strings (CoS) parameter $\alpha$ and the quantum deformation parameter $a$. Other physical parameters are fixed as: BH mass $M=1$, cosmological constant $\Lambda = -0.01$, normalization constant of the quintessence field $\mathrm{N} = 0.01$ with an equation-of-state parameter $w = -2/3$. The results are shown for two different electric charge values, namely $Q = 0.5$ and $Q = 1$.

\section{Optical Properties of BH} \label{S3}

In this section, we study null geodesic motion, deriving the photon trajectories, the photon sphere and BH shadows. We demonstrate how geometric and physical parameters influence the optical properties. Recent studies of optical properties in various BH solution under different configurations and topological features have reported in \cite{FA1,FA2,FA3,FA4,FA5,FA6,FA7,FA8,FA9,FA10,FA11,FA12} and related references therein.

Considering the geodesic motion in the equatorial plane, defined by $\theta=\pi/2$ and $\dot{\theta}=0$. The Lagrangian density function is given by, $\mathcal{L}=\frac{1}{2}\,g_{\mu\nu}\,\frac{dx^{\mu}}{d\lambda}\,\frac{dx^{\nu}}{d\lambda}$, where $\lambda$ is an affine parameter along the geodesic paths, $g_{\mu\nu}$ is the metric tensor.

Expressing the space-time (\ref{aa4}) in the form $ds^2=g_{\mu\nu}\,dx^{\mu}\,dx^{\nu}$ ($\mu,\nu=0,1,2,3$), the Lagrangian density function can be expressed as, 
\begin{equation}
    \mathcal{L}=\frac{1}{2}\,\left[-f(r)\,\left(\frac{dt}{d\lambda}\right)^2+\frac{1}{f(r)}\,\left(\frac{dr}{d\lambda}\right)^2+r^2\,\left(\frac{d\phi}{d\lambda}\right)^2\right].\label{bb1}
\end{equation}

One can see that the Lagrangian density function is independent of the temporal coordinate $t$ and the azimuthal coordinate $\phi$. Therefore, there exist two conserved quantities associated with these cyclic coordinates and are given by
\begin{equation}
    \mathrm{E}=f(r)\,\frac{dt}{d\lambda}.\label{bb1a}
\end{equation}
And
\begin{equation}
    \mathrm{L}=r^2\,\frac{d\phi}{d\lambda}.\label{bb2}
\end{equation}
Here $\mathrm{E}$ is the conserved energy and $\mathrm{L}$ is the conserved angular momentum.

\begin{figure}[ht!]
    \centering
    \includegraphics[width=0.3\linewidth]{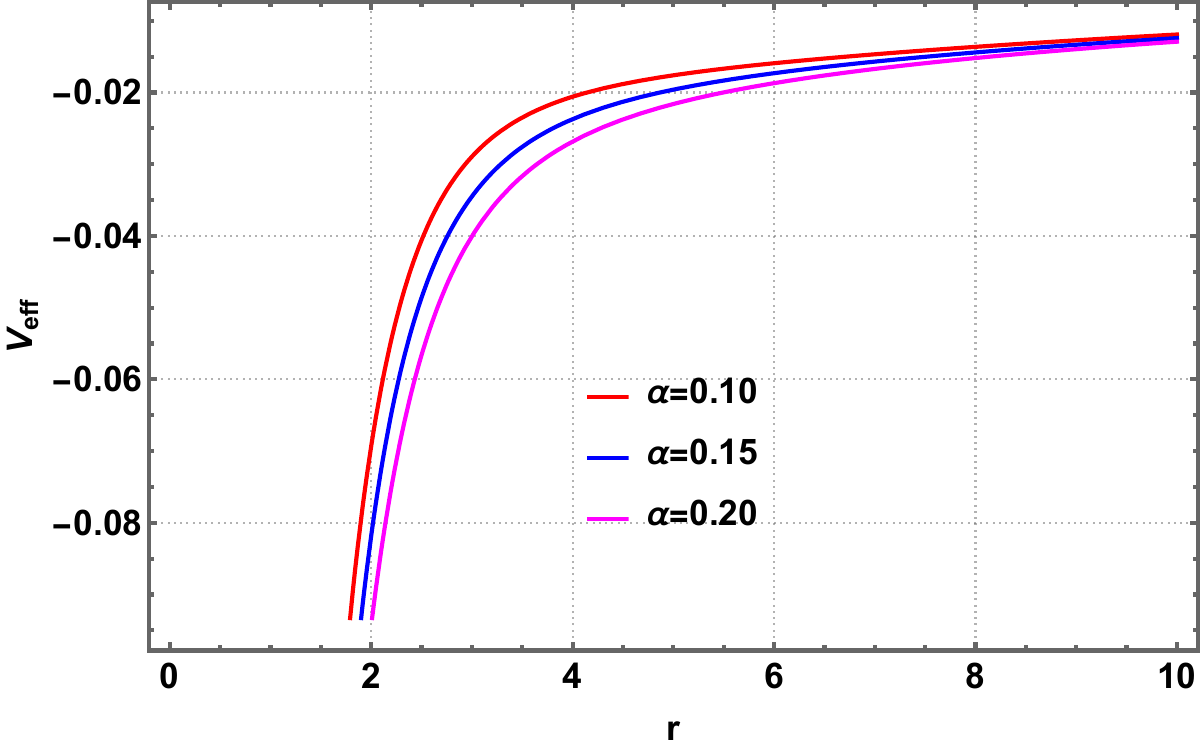}\quad
    \includegraphics[width=0.3\linewidth]{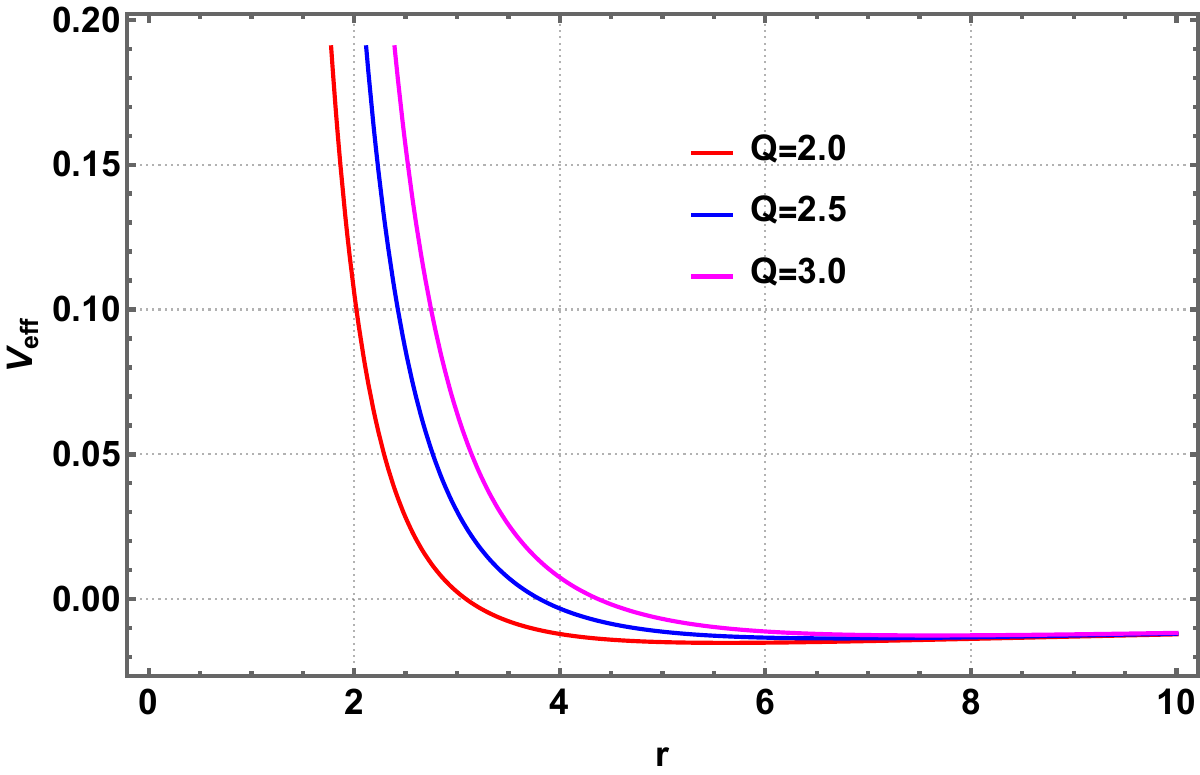}\quad
    \includegraphics[width=0.3\linewidth]{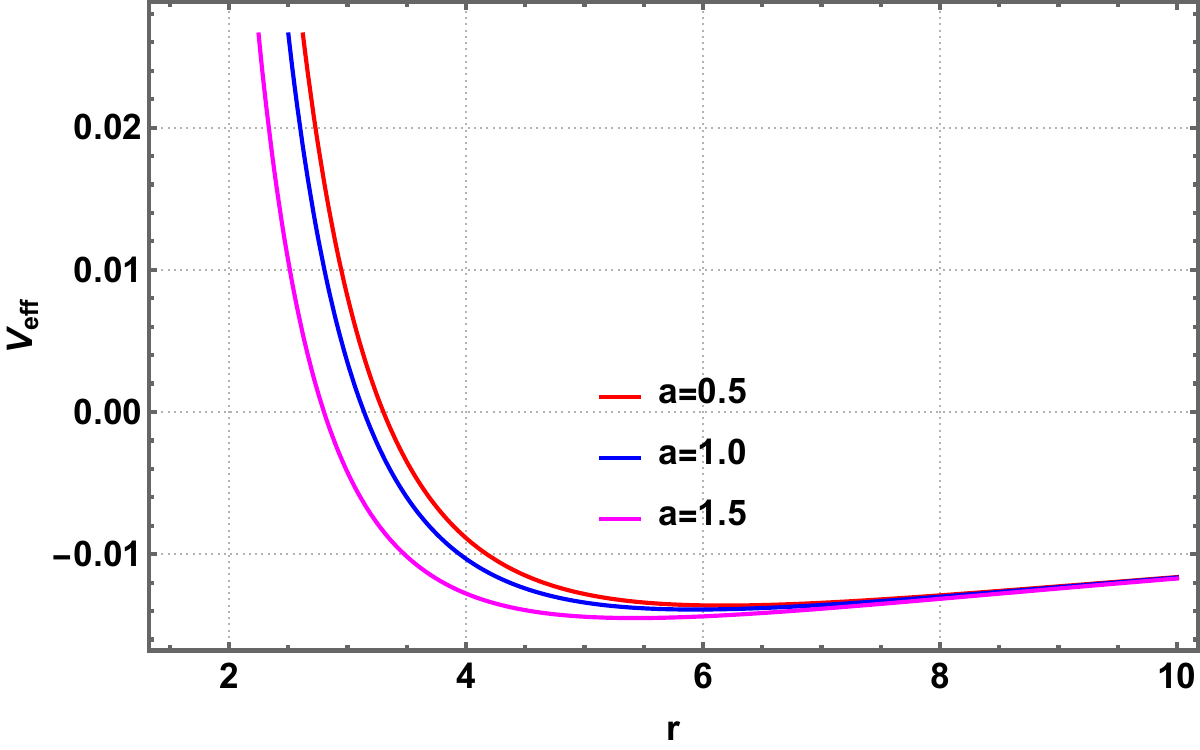}\\
    (a) $Q=1,\, a=0.5$ \hspace{4cm}  (b) $\alpha=0.15,\, a=0.5$ \hspace{4cm} (c) $\alpha=0.1, \, Q=2$
    \caption{\footnotesize Behavior of the effective potential $V_\text{eff}(r)$ given in Eq.~(\ref{bb4}) by varying values of CS parameter $\alpha$, electric charge $Q$, and quantum deformation parameter $a$. Here $M=1, w=-2/3, \mathrm{L}=1, \mathrm{N}=0.2, \Lambda=-0.003$.}
    \label{fig:null}
\end{figure}

Eliminating $\frac{dt}{d\lambda}$ and $\frac{d\phi}{d\lambda}$ using Eqs.~(\ref{bb1a}) and (\ref{bb2}) into the Eq. (\ref{bb1}), we find the following equation for null geodesics motion as,
\begin{equation}
    \left(\frac{dr}{d\lambda}\right)^2+V_\text{eff}(r)=\mathrm{E}^2\label{bb3}
\end{equation}
which is equivalent to a one-dimensional equation of motion of a particle of unit mass having energy $\mathrm{E}^2$ and the potential $V_\text{eff}(r)$. This potential also called the effective potential of null geodesic system governs the dynamics of the photon particles in the gravitational field of BH solution. In our case, this effective potential is given by
\begin{equation}
    V_\text{eff}(r)=\frac{\mathrm{L}^2}{r^2}\,f(r)=\frac{\mathrm{L}^2}{r^2}\,\left[1-\alpha-\frac{2\,M}{r}+\frac{(Q^2-a/2)}{r^2}-\frac{\mathrm{N}}{r^{3\,\omega+1}}-\frac{\Lambda}{3}\,r^2\right].\label{bb4}
\end{equation}

From the expression (\ref{bb4}), we observe that the effective potential is influenced by geometric and physical parameters. These include the string cloud characterized by the parameter $\alpha$, the quantum deformation characterized by the parameter $a$, the electric charge $Q$, the quintessence-like fluid characterized by the parameters $\mathrm{N}$ and $w$. Additionally, the BH mass $M$ and the cosmological constant $\Lambda$ modifies this effective potential. Collectively, these parameters shape the structure of the effective potential and significantly influence the dynamics of null geodesics.

\begin{center}
    {\bf I.\,\, Photon Trajectories}
\end{center}

In this part, we discuss trajectories of light and show how geometric and physical parameters alter the trajectory of light in the given gravitational field.

The equation of orbit using Eqs.~(\ref{bb2}) and (\ref{bb3}) is defined by
\begin{equation}
    \left(\frac{1}{r^2}\,\frac{dr}{d\phi}\right)^2+\frac{(1-\alpha)}{r^2}=\frac{1}{\beta^2}+\frac{\Lambda}{3}+\frac{2\,M}{r^3}-\frac{(Q^2-a/2)}{r^4}+\frac{\mathrm{N}}{r^{3\,\omega+3}}.\label{bb6}
\end{equation}
Transforming to a new variable via $r=\frac{1}{u}$ into the above equation yields:
\begin{equation}
    \left(\frac{du}{d\phi}\right)^2+(1-\alpha)\,u^2=\frac{1}{\beta^2}+\frac{\Lambda}{3}+2\,M\,u^3-(Q^2-a/2)\,u^4+\mathrm{N}\,u^{3\,w+3}.\label{bb7}
\end{equation}
Differentiating both sides of Eq.~(\ref{bb7}) w. r. to $\phi$ and after simplification results:
\begin{equation}
    \frac{d^2u}{d\phi^2}+(1-\alpha)\,u=3\,M\,u^2-(2\,Q^2-a)\,u^3+\frac{1}{2}\,\mathrm{N}\,(3\,w+3)\,u^{3\,w+2}.\label{bb8}
\end{equation}

\begin{figure}[ht!]
    \centering
    \includegraphics[width=0.26\linewidth]{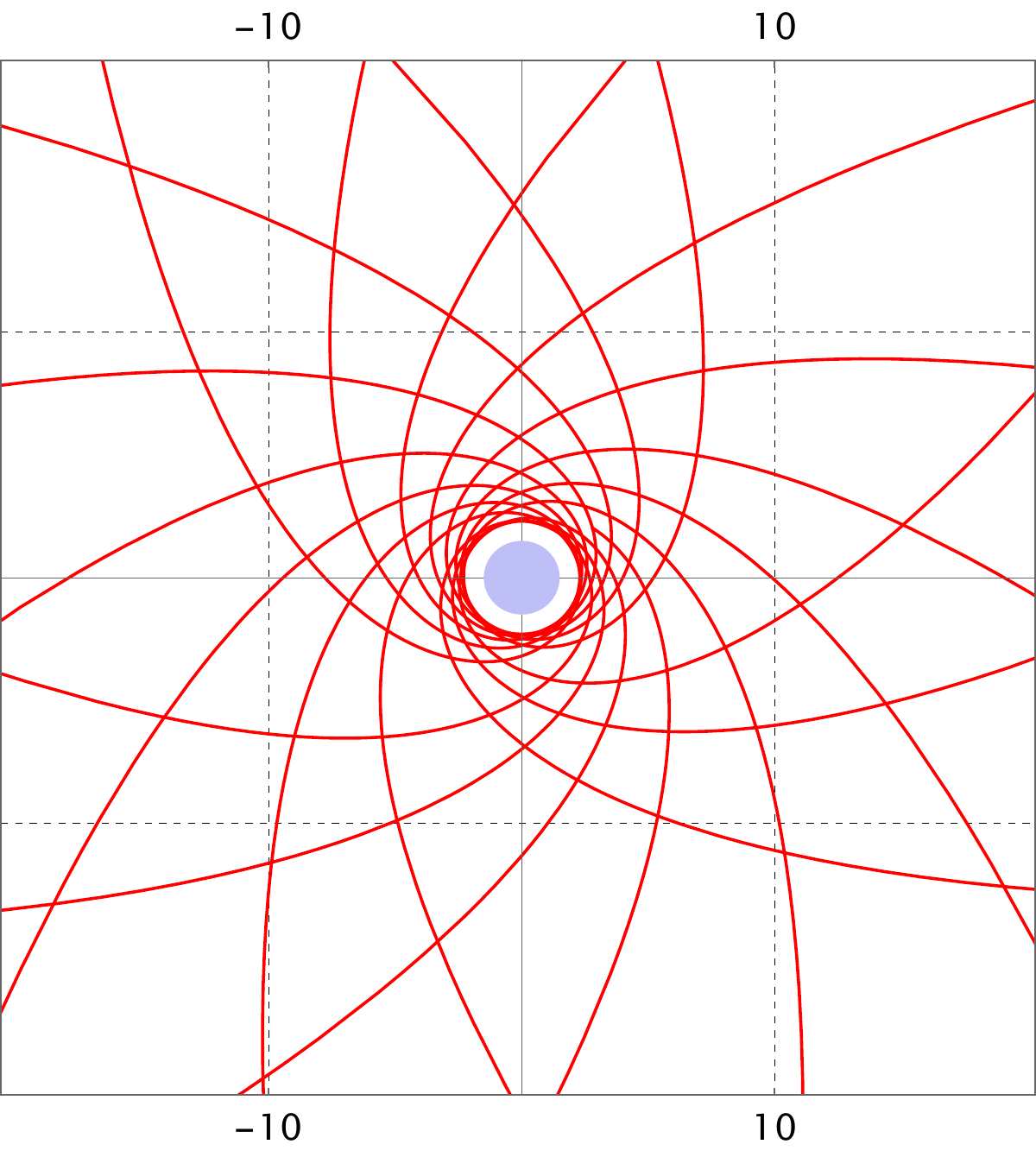}\quad\quad
    \includegraphics[width=0.26\linewidth]{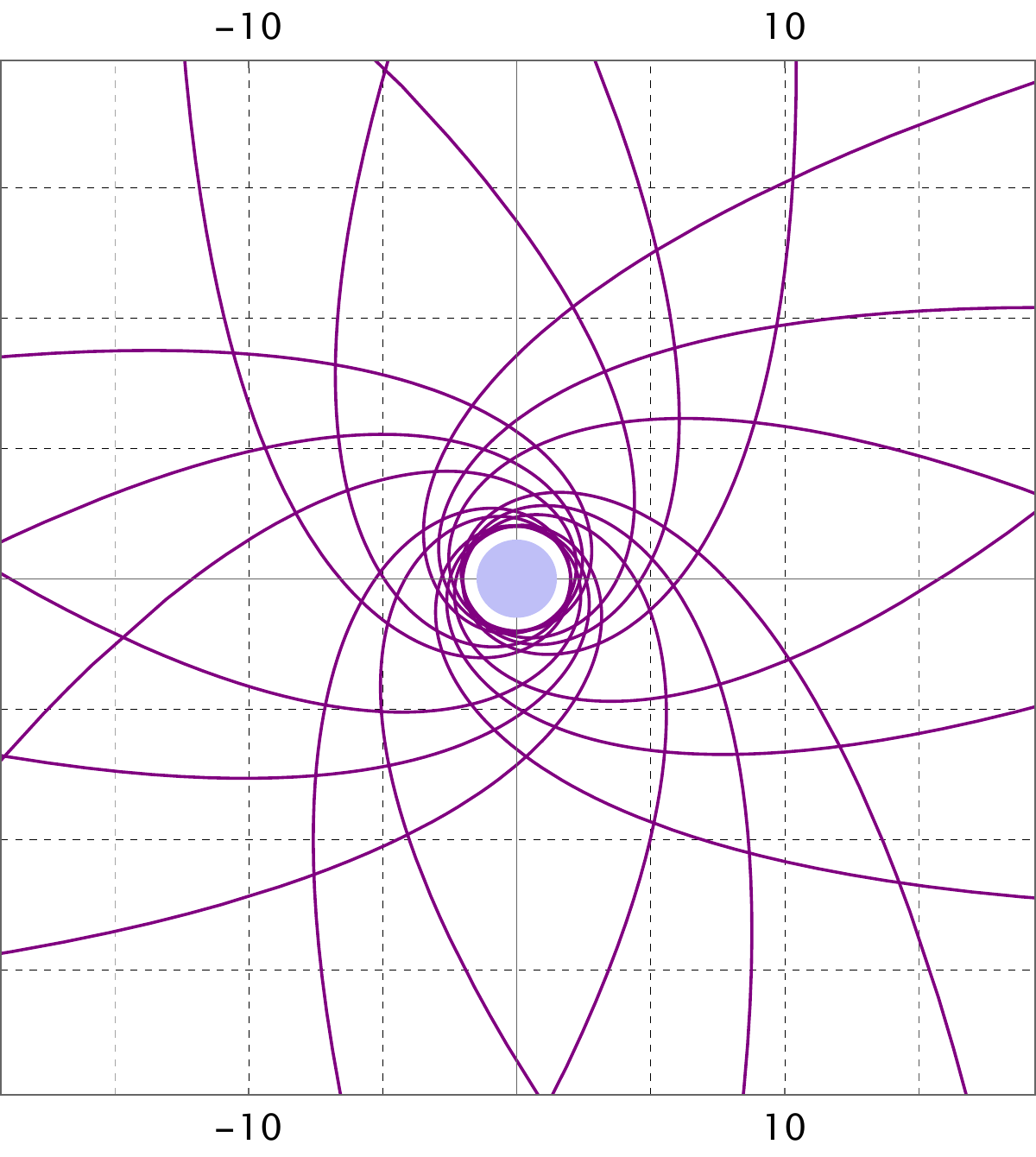}\quad\quad
    \includegraphics[width=0.26\linewidth]{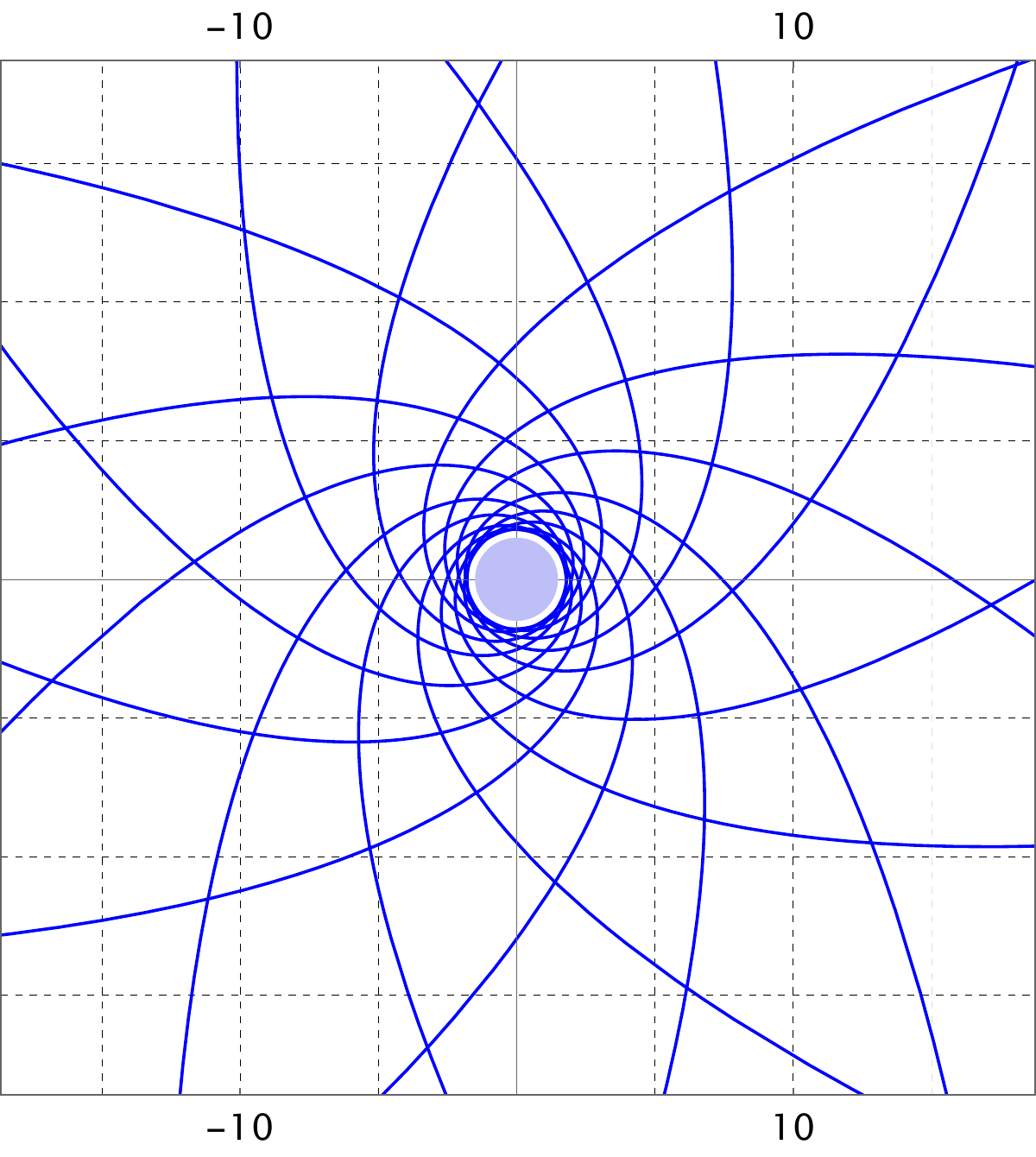}\\
    (a) $\alpha=0.05$ \hspace{5cm} (b) $\alpha=0.10$ \hspace{5cm} (c) $\alpha=0.15$
    \caption{\footnotesize Parametric plot of light trajectory $r(\phi)=\frac{1}{u(\phi)}$ for different values of CS parameter $\alpha$. Here, we set $M=1, \mathrm{N}=0.2, Q=1.5, a=0.5$. The initial boundary conditions are $u(0)=0.001, u'(0)=0.01$.}
    \label{fig:trajectory-1}
    \includegraphics[width=0.26\linewidth]{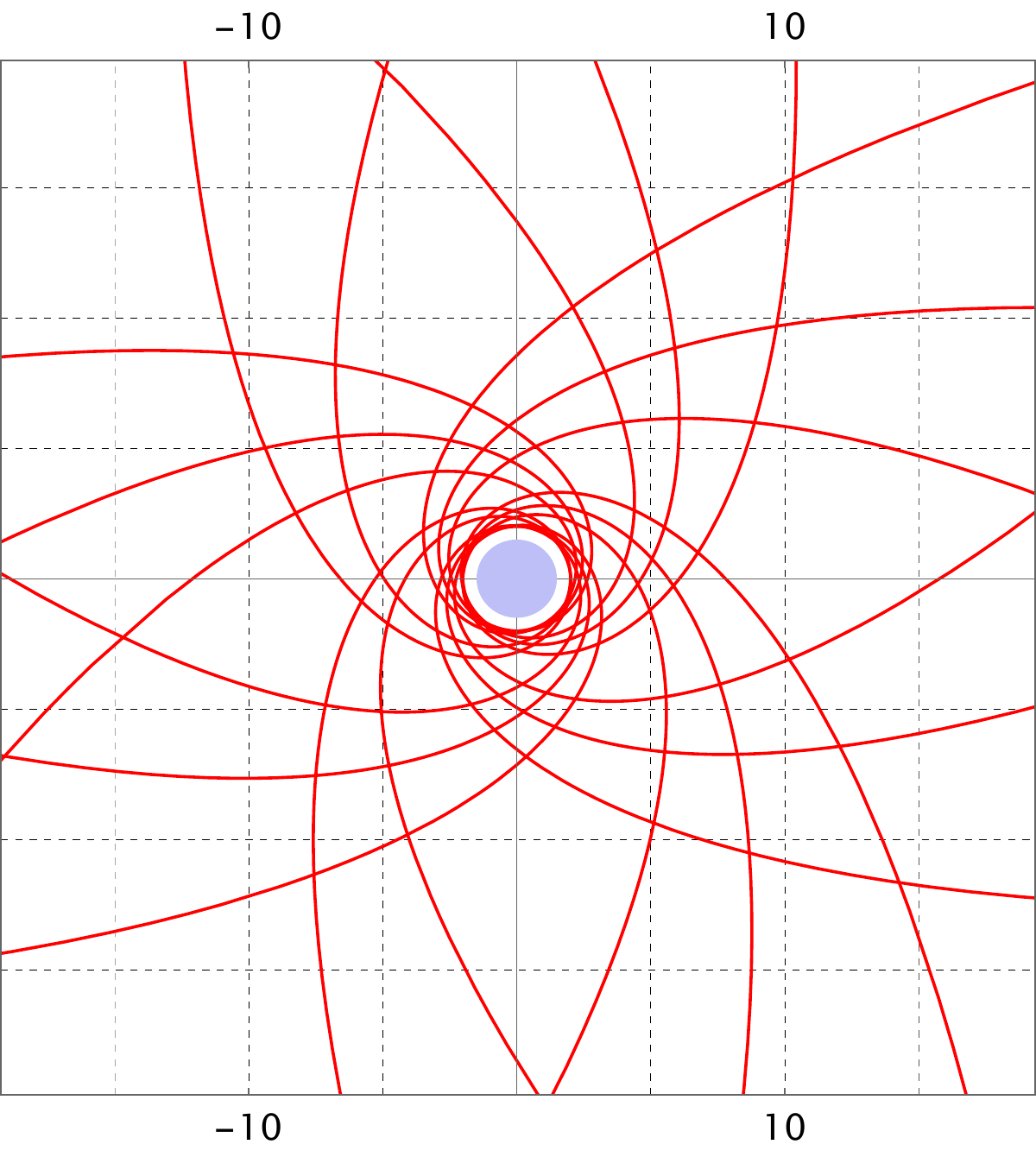}\quad\quad
    \includegraphics[width=0.26\linewidth]{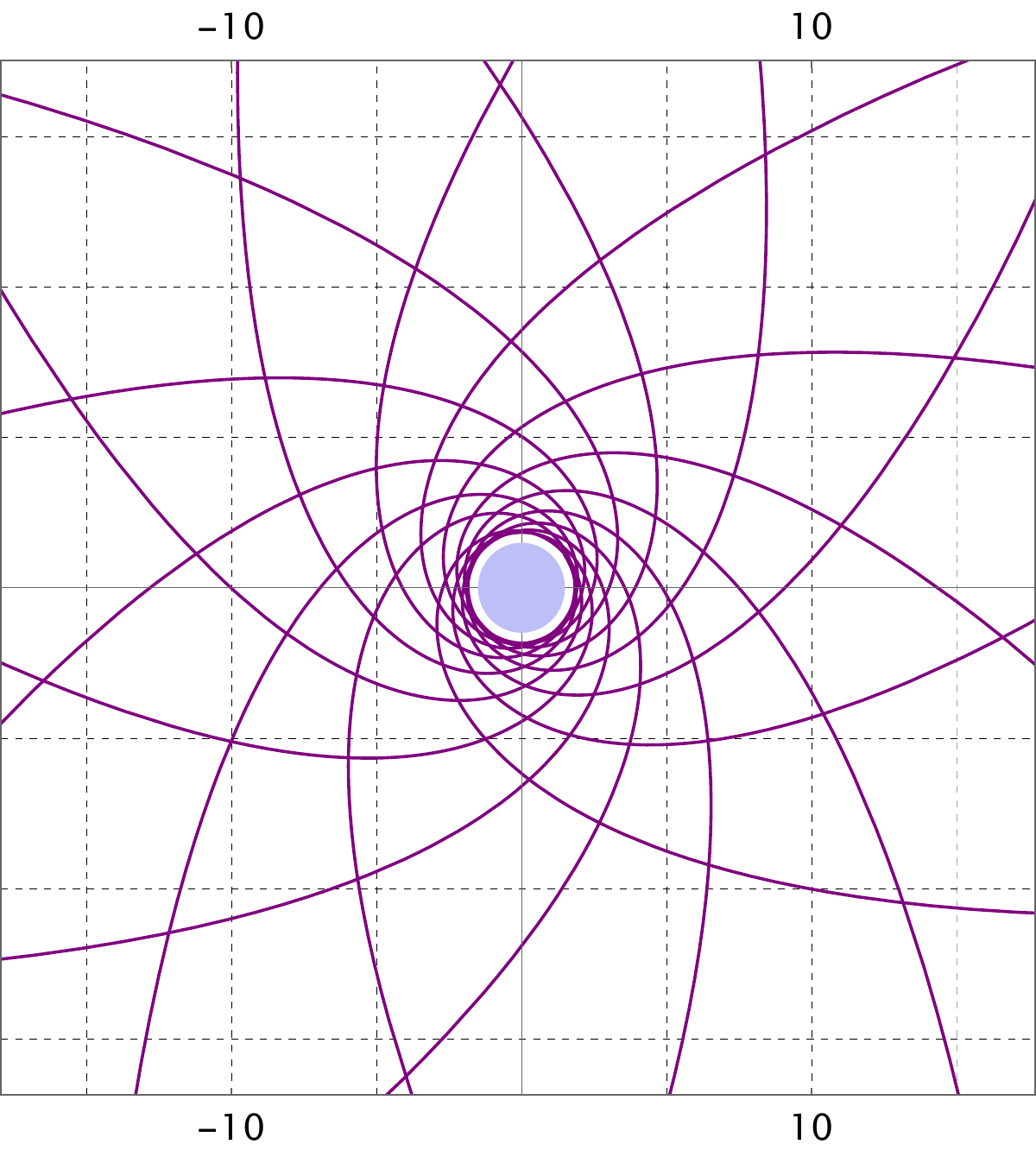}\quad\quad
    \includegraphics[width=0.26\linewidth]{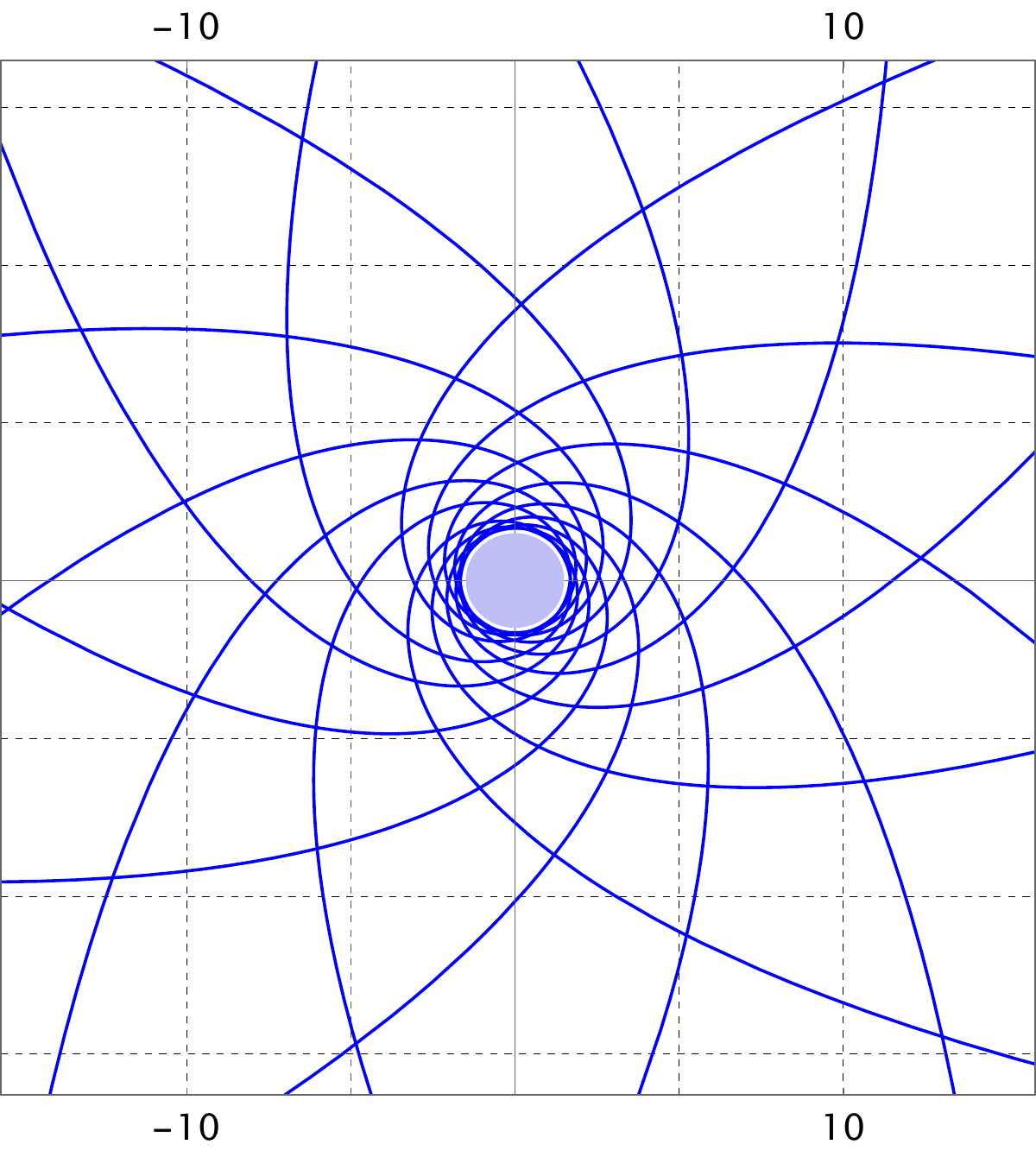}\\
    (a) $a=0.50$ \hspace{5cm} (b) $a=0.75$ \hspace{5cm} (c) $a=1.00$
    \caption{\footnotesize Parametric plot of light trajectory $r(\phi)=\frac{1}{u(\phi)}$ for different values of the quantum deformation parameter $a$. Here, we set $M=1, \mathrm{N}=0.2, Q=1.5, \alpha=0.1$. The initial boundary conditions are $u(0)=0.001, u'(0)=0.01$.}
    \label{fig:trajectory-2}
    \includegraphics[width=0.26\linewidth]{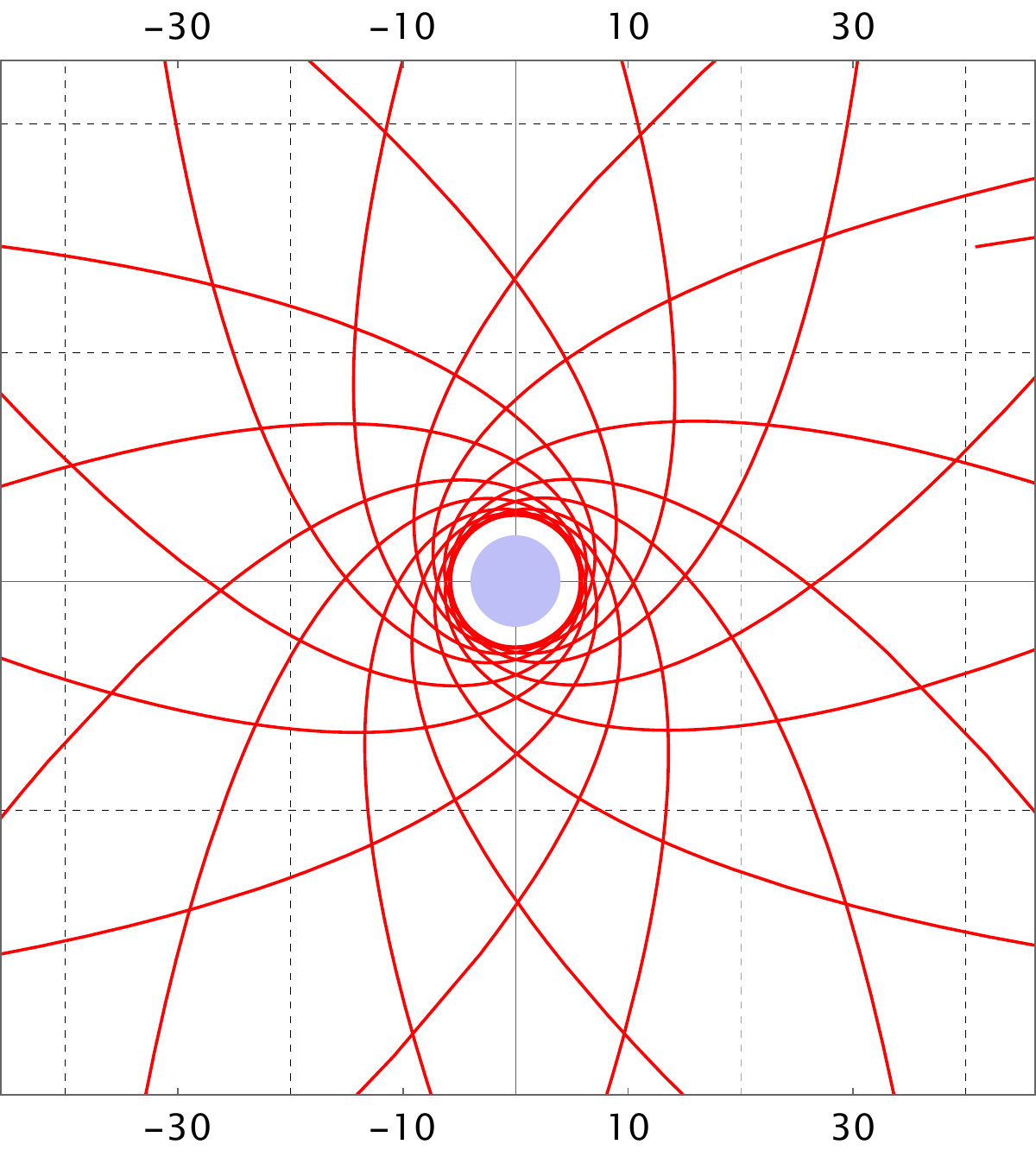}\quad\quad
    \includegraphics[width=0.26\linewidth]{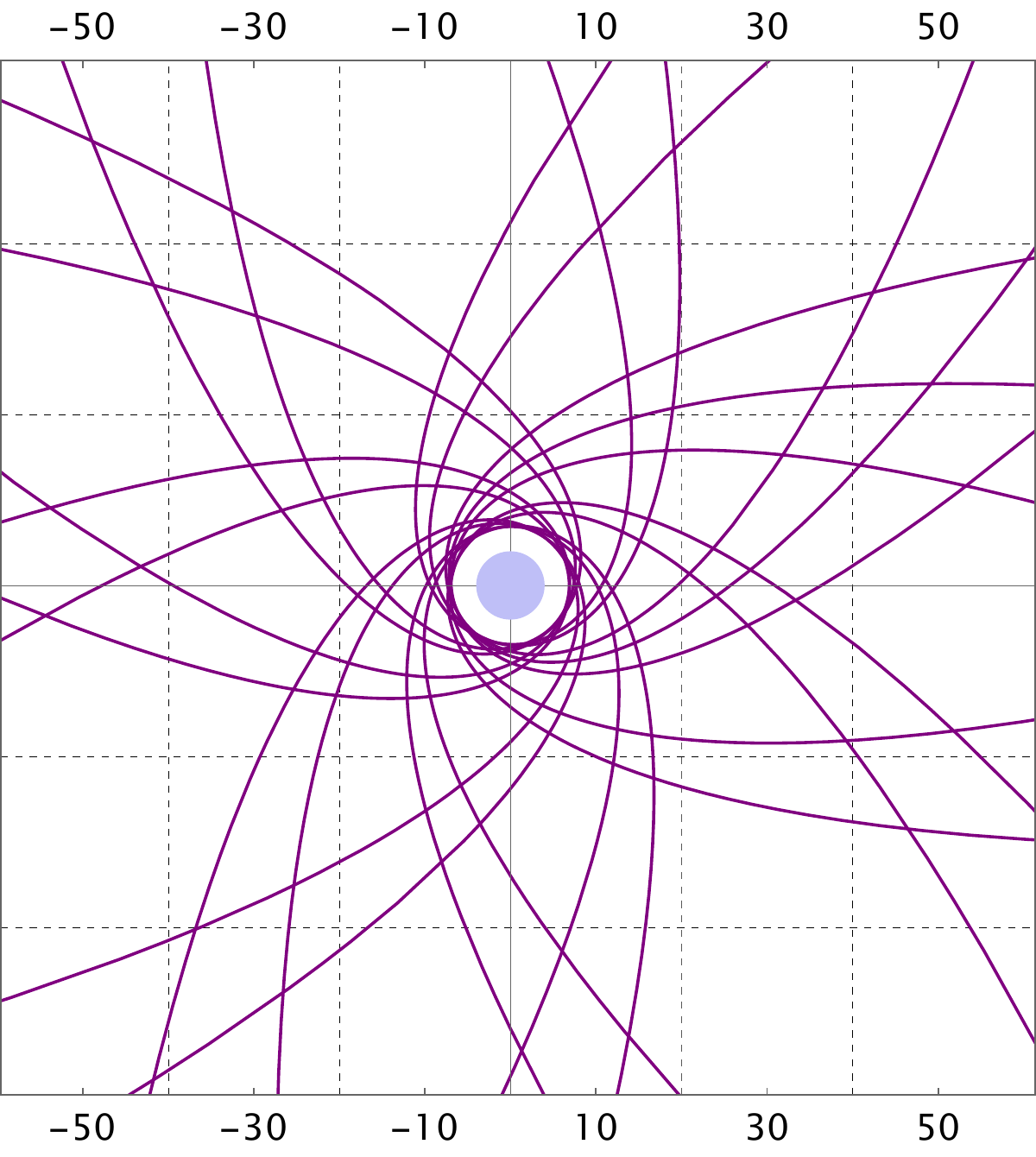}\quad\quad
    \includegraphics[width=0.26\linewidth]{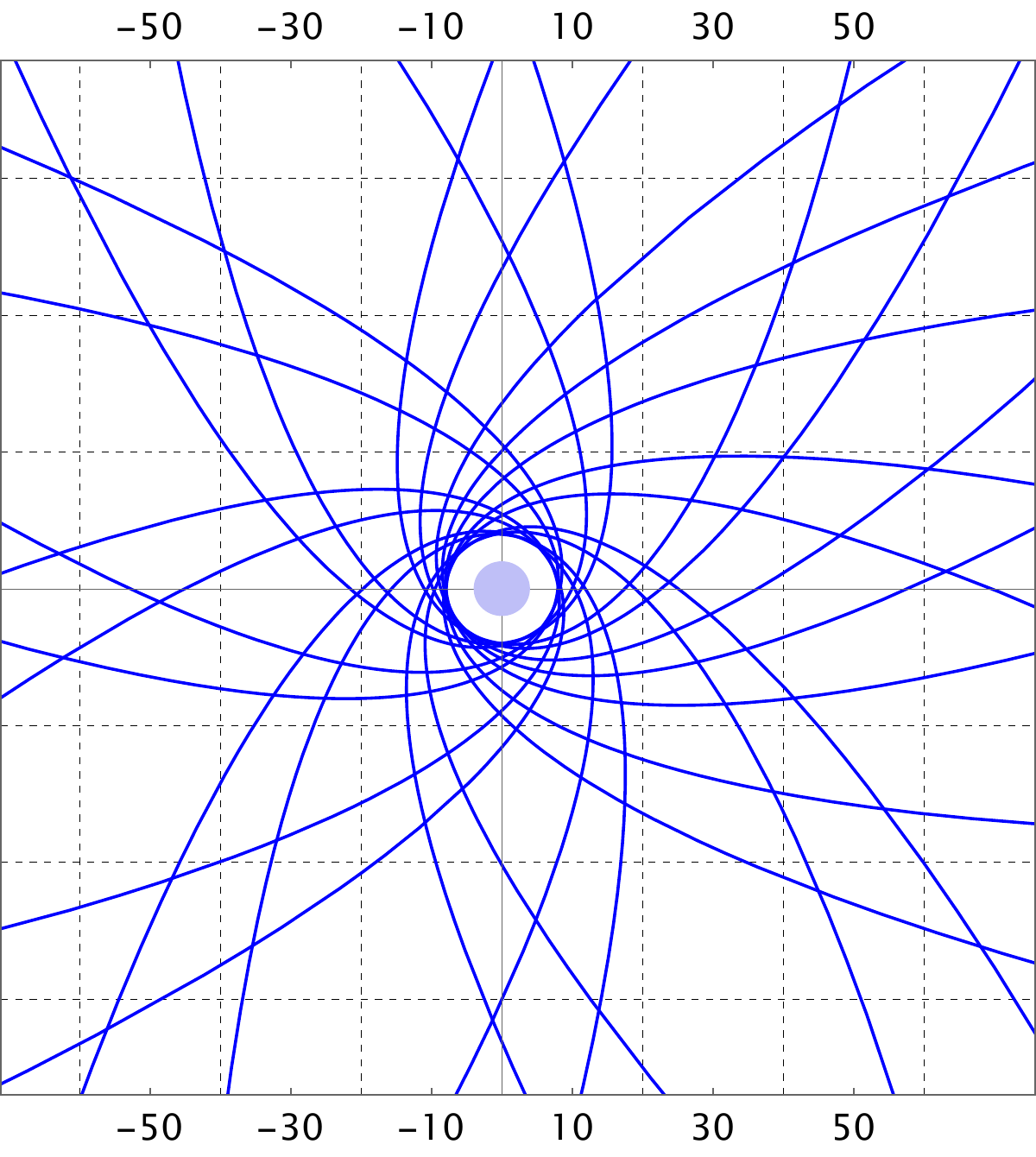}\\
    (a) $Q=1$ \hspace{5cm} (b) $Q=2$ \hspace{5cm} (c) $Q=3$
    \caption{\footnotesize Parametric plot of light trajectory $r(\phi)=\frac{1}{u(\phi)}$ for different values of the electric charge $Q$. Here, we set $M=1, \mathrm{N}=0.1, \alpha=0.1, a=0.5$. The initial boundary conditions are $u(0)=0.001, u'(0)=0.01$.}
    \label{fig:trajectory-3}
\end{figure}

Equation (\ref{bb8}) is a nonlinear, second-order partial differential equation that describes the trajectory of light in the given gravitational field. This trajectory is influenced by several parameters, including the string cloud characterized by the parameter $\alpha$, the quantum deformation characterized by the parameter $a$, the electric charge $Q$, the quintessence-like fluid characterized by the parameters $\mathrm{N}$ and $w$. Additionally, the BH mass $M$ modifies this trajectory.

For a particular state parameter, $w=-2/3$, this trajectory equation reduces as,
\begin{equation}
    \frac{d^2u}{d\phi^2}+(1-\alpha)\,u=3\,M\,u^2-(2\,Q^2-a)\,u^3+\frac{1}{2}\,\mathrm{N}.\label{bb9}
\end{equation}

\begin{center}
    \large{\bf II.\,\,Circular orbits, photon sphere and BH shadows}
\end{center}

In this part, we study the photon sphere radius, stable or unstable circular null orbits and finally BH shadows using the effective potential $V_\text{eff}(r)$ derived earlier given in Eq. (\ref{bb4}). We show how geometric and physical parameters influences the optical properties and alter these.

For circular orbits of radius $r=r_c$, the conditions $\frac{dr}{d\lambda}=0$ and $\frac{d^2r}{d\lambda^2}=0$ must be satisfied. These condition using Eq. (\ref{bb3}) yields
\begin{equation}
    \mathrm{E}^2_\text{ph}=V_\text{eff}(r).\label{bb10}
\end{equation}
And
\begin{equation}
    V'_\text{eff}(r)=0.\label{bb10a}
\end{equation}

Using the effective potential given in Eq. (\ref{bb4}) into the relation (\ref{bb10}), we find the critical impact parameter for light at radius $r=r_c$ as,
\begin{equation}
    \beta_c=\frac{r_c}{\sqrt{1-\alpha-\frac{2\,M}{r_c}+\frac{(Q^2-a/2)}{r^2_c}-\frac{\mathrm{N}}{r^{3\,\omega+1}_c}-\frac{\Lambda}{3}\,r^2_c}}.\label{bb11}
\end{equation}
This parameter is influenced by the string cloud parameter $\alpha$, the quantum deformation parameter $a$, the electric charge $Q$, the normalization constant $\mathrm{N}$ and state parameter $w$ of the quintessence-like fluid. Additionally, the BH mass $M$ and the cosmological constant $\Lambda$ modifies this impact parameter.

For a particular state parameter, $w=-2/3$, the critical impact parameter reduces as,
\begin{equation}
    \beta_c=\frac{r_c}{\sqrt{1-\alpha-\frac{2\,M}{r_c}+\frac{(Q^2-a/2)}{r^2_c}-\mathrm{N}\,r_c-\frac{\Lambda}{3}\,r^2_c}}.\label{bb12}
\end{equation}

Using the effective potential given in Eq.~(\ref{bb4}) into the equation (\ref{bb10a}), we find the photon sphere radius $r=r_\text{ph}$ satisfying the following relation
\begin{equation}
    \mathrm{N}\,(3\,w+3)\,r^3-2\,(1-\alpha)\,r^{3\,\omega+4}+6\,M\,r^{3\,\omega+3}-2\,(2\,Q^2-a)\,r^{3\,\omega+2}=0.\label{bb13}
\end{equation}
From the above equation (\ref{bb13}), we observe that the string cloud parameter $\alpha$, quantum deformation parameter $a$,  electric charge $Q$, normalization constant $\mathrm{N}$ and state parameter $w$ of the quintessence-like fluid, and BH mass $M$ all together effects the radius of the photon sphere. 

This equation (\ref{bb13}) in general can not be solved because it is a polynomial equation of $r$ of degree $(3\,w+1)$. To obtain a simple equation, we set the state parameter $w=-2/3$. This setting simplifies the above equation as,
\begin{equation}
    \mathrm{N}\,r^3-2\,(1-\alpha)\,r^2+6\,M\,r-2\,(2\,Q^2-a)=0.\label{bb14}
\end{equation}
The above equation is a cubic polynomial in $r=r_\text{ph}$, whose real-valued solution corresponds to the radius of the photon sphere. Although an analytical solution can be obtained using standard methods for solving cubic equations, it tends to be algebraically complicated. Alternatively, a numerical solution can be efficiently computed by assigning suitable values to the relevant geometric and physical parameters.

Next, we focus into the BH shadows cast by the BH solution, and examine how the geometric and physical parameters influence the size of the shadow. The shadow size $ R_s $ of a spherically symmetric BH (as seen by a distant observer) is determined by the photon sphere radius $ r_\text{ph} $ and is typically expressed using the critical impact parameter $ \beta_c $. The radius of the shadow can be determined using the following formula:
\begin{equation}
R_s=\beta_c=\frac{r_\text{ph}}{\sqrt{f(r_{ph})}}.\label{bb15}    
\end{equation}
Using the metric function Eq.~(\ref{aa6}), the shadow size can be expressioned as,
\begin{equation}
R_s=\frac{r_\text{ph}}{\sqrt{1-\alpha-\frac{2\,M}{r_\text{ph}}+\frac{(Q^2-a/2)}{r^2_\text{ph}}-\frac{\mathrm{N}}{r^{3\,w+1}_\text{ph}}-\frac{\Lambda}{3}\,r^2_\text{ph}}}.\label{bb16}
\end{equation}

From the above expression (\ref{bb15}), we observe that the size of BH shadow depends on the string cloud parameter $\alpha$, quantum deformation parameter $a$,  electric charge $Q$, normalization constant $\mathrm{N}$ and state parameter $w$ of the quintessence-like fluid, the BH mass $M$, and the cosmological constant $\Lambda$, all together effects the size of the BH shadows. 

For a particular state parameter, $w=-2/3$, the shadow size reduces as,
\begin{equation}
R_s=\frac{r_\text{ph}}{\sqrt{1-\alpha-\frac{2\,M}{r_\text{ph}}+\frac{(Q^2-a/2)}{r^2_\text{ph}}-\mathrm{N}\,r_\text{ph}-\frac{\Lambda}{3}\,r^2_\text{ph}}}.\label{bb17}
\end{equation}

\begin{table}[ht!]
\centering
\begin{tabular}{|c||c c|c c|c c|c c|c c|}
\hline
\multirow{2}{*}{$a$} & \multicolumn{2}{c|}{$\alpha = 0.05$} & \multicolumn{2}{c|}{$\alpha = 0.10$} & \multicolumn{2}{c|}{$\alpha = 0.15$} & \multicolumn{2}{c|}{$\alpha = 0.20$} & \multicolumn{2}{c|}{$\alpha = 0.25$} \\
\cline{2-11}
& $r_{\text{ph}}$ & $R_s$ & $r_{\text{ph}}$ & $R_s$ & $r_{\text{ph}}$ & $R_s$ & $r_{\text{ph}}$ & $R_s$ & $r_{\text{ph}}$ & $R_s$ \\
\hline
0.50  & 3.3955 & 6.3234 & 3.5927 & 6.9162 & 3.8164 & 7.6129 & 4.0724 & 8.4418 & 4.3689 & 9.4419 \\
0.75  & 3.4743 & 6.4401 & 3.6723 & 7.0381 & 3.8967 & 7.7407 & 4.1538 & 8.5763 & 4.4514 & 9.5841 \\
1.00  & 3.5500 & 6.5527 & 3.7488 & 7.1559 & 3.9743 & 7.8645 & 4.2323 & 8.7068 & 4.5312 & 9.7223 \\
1.25  & 3.6229 & 6.6616 & 3.8227 & 7.2701 & 4.0492 & 7.9846 & 4.3084 & 8.8336 & 4.6087 & 9.8567 \\
1.50  & 3.6934 & 6.7673 & 3.8942 & 7.3810 & 4.1217 & 8.1013 & 4.3823 & 8.9570 & 4.6839 & 9.9877 \\
\hline
\end{tabular}
\caption{\footnotesize Photon sphere radius $ r_{\text{ph}} $ and shadow radius $ R_s $ for various values of $ a $ and $ \alpha $ with $Q=0.25$. Here, we set $M=1, \mathrm{N}=0.02, \Lambda=-0.003$}
\label{tab:shadow-1}
\end{table}

\begin{table}[ht!]
\centering
\begin{tabular}{|c||c c|c c|c c|c c|c c|}
\hline
\multirow{2}{*}{$a$} & \multicolumn{2}{c|}{$\alpha = 0.05$} & \multicolumn{2}{c|}{$\alpha = 0.10$} & \multicolumn{2}{c|}{$\alpha = 0.15$} & \multicolumn{2}{c|}{$\alpha = 0.20$} & \multicolumn{2}{c|}{$\alpha = 0.25$} \\
\cline{2-11}
& $r_{\text{ph}}$ & $R_s$ & $r_{\text{ph}}$ & $R_s$ & $r_{\text{ph}}$ & $R_s$ & $r_{\text{ph}}$ & $R_s$ & $r_{\text{ph}}$ & $R_s$ \\
\hline
0.50  & 3.2705 & 6.1396 & 3.4669 & 6.7246 & 3.6896 & 7.4125 & 3.9445 & 8.2314 & 4.2397 & 9.2200 \\
\hline
0.75  & 3.3548 & 6.2634 & 3.5517 & 6.8535 & 3.7750 & 7.5473 & 4.0306 & 8.3729 & 4.3266 & 9.3691 \\
\hline
1.00  & 3.4353 & 6.3823 & 3.6329 & 6.9777 & 3.8569 & 7.6773 & 4.1134 & 8.5096 & 4.4105 & 9.5135 \\
\hline
1.25  & 3.5125 & 6.4969 & 3.7109 & 7.0975 & 3.9358 & 7.8031 & 4.1934 & 8.6420 & 4.4916 & 9.6537 \\
\hline
1.50  & 3.5868 & 6.6076 & 3.7861 & 7.2135 & 4.0120 & 7.9249 & 4.2707 & 8.7706 & 4.5702 & 9.7899 \\
\hline
\end{tabular}
\caption{\footnotesize Photon sphere radius $ r_{\text{ph}} $ and shadow radius $ R_s $ for various values of $ a $ and $ \alpha $ with $Q=0.50$. Here, we set $M=1, \mathrm{N}=0.02, \Lambda=-0.003$.}
\label{tab:shadow-2}
\end{table}

Finally, we focus on the stability or instability of circular null orbits and demonstrate how geometric and physical parameters influence this behavior. To analyze stability, we use the Lyapunov exponent, a key tool for characterizing the nature of these orbits. For a static and spherically symmetric space-time, the Lyapunov exponent is given by \cite{VC}
\begin{equation}
    \lambda_L=\sqrt{-\frac{1}{2}\,\frac{V''_\text{eff}(r)}{\dot{t}^2}}.\label{bb18}
\end{equation}
Using the effective potential given in Eq.~(\ref{bb4}), we find the following expression for the Lyapunov exponent as,
\begin{equation}
    \lambda_L=\frac{1}{r}\,\sqrt{1 - \alpha- \frac{(2\,Q^2-a)}{r^2}
+ \frac{N \left[(3\omega + 1)(3\omega + 2) - 2\right]}{2\,r^{3\,\omega + 1}}}\,\sqrt{1-\alpha-\frac{2\,M}{r}+\frac{(Q^2-a/2)}{r^2}-\frac{\mathrm{N}}{r^{3\,\omega+1}}-\frac{\Lambda}{3}\,r^2}.\label{bb19}
\end{equation}

This Lyapunov exponent (\ref{bb19}) is influenced by the string cloud parameter $\alpha$, the quantum deformation parameter $a$, the electric charge $Q$, the normalization constant $\mathrm{N}$ and state parameter $w$ of the quintessence-like fluid. Additionally, the BH mass $M$ and the cosmological constant $\Lambda$ modifies this. 

Using this expression, one can not determined whether the circular null orbits are stable or unstable since state parameter $w$ is not set yet. Let us consider one particular value of the state parameter $w=-2/3$ and check this criteria of circular orbits. Thus, for the state parameter $w=-2/3$, this Lyapunov exponent from Eq. (\ref{bb19}) reduces as,
\begin{equation}
    \lambda_L=\frac{1}{r}\,\sqrt{1 - \alpha- \frac{(2\,Q^2-a)}{r^2}
-\mathrm{N}\,r}\,\sqrt{1-\alpha-\frac{2\,M}{r}+\frac{(Q^2-a/2)}{r^2}-\mathrm{N}\,r-\frac{\Lambda}{3}\,r^2}.\label{bb20}
\end{equation}

It is well-known in BH physics that 
\begin{itemize}
    \item \textbf{Positive Lyapunov Exponent} ($\lambda_L > 0$): A positive exponent signifies that perturbations grow exponentially with time, indicating that the circular null orbit is \emph{unstable}.

    \item \textbf{Zero Lyapunov Exponent} ($\lambda_L = 0$): When the exponent is zero, the orbit is \emph{marginally stable}. In this critical state, perturbations neither exponentially grow nor decay but may persist or evolve slowly over time. This marginal stability marks a threshold between stable and unstable regimes and often corresponds to extremal.

    \item \textbf{Imaginary Lyapunov Exponent} ($\lambda_L =\omega_R+ i\, \omega_I$, $\omega_R, \omega_I \in \mathbb{R}$): An imaginary exponent implies that perturbations exhibit \emph{bounded oscillatory motion} around the circular null orbit, indicating \emph{stability}. Photons near such stable orbits remain confined, oscillating about the equilibrium radius rather than escaping.
\end{itemize}

Therefore, analyzing the sign and nature of the Lyapunov exponent not only shows the stability characteristics of photon spheres but also provides insight into the optical properties of BHs, influencing phenomena such as gravitational lensing, quasinormal mode spectra, and the morphology of BH shadows observed by instruments like the Event Horizon Telescope.

\begin{figure}[ht!]
    \centering
    \includegraphics[width=0.35\linewidth]{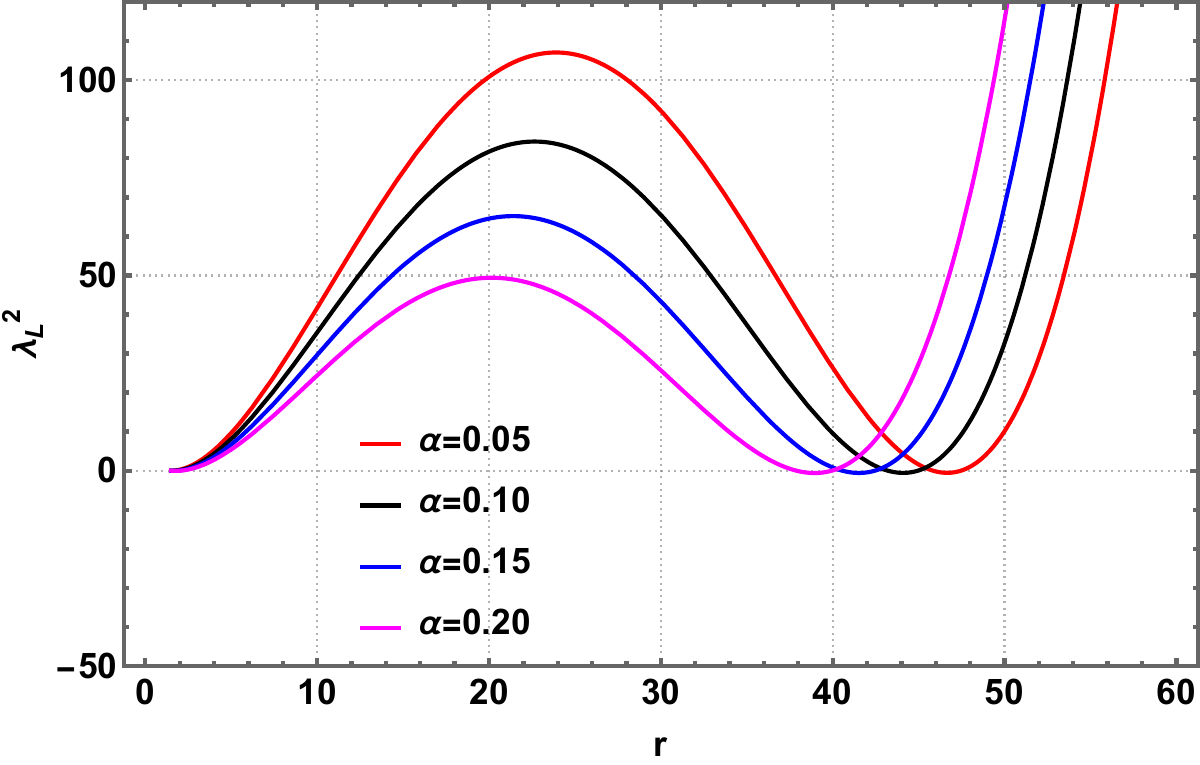}\quad\quad\quad
    \includegraphics[width=0.35\linewidth]{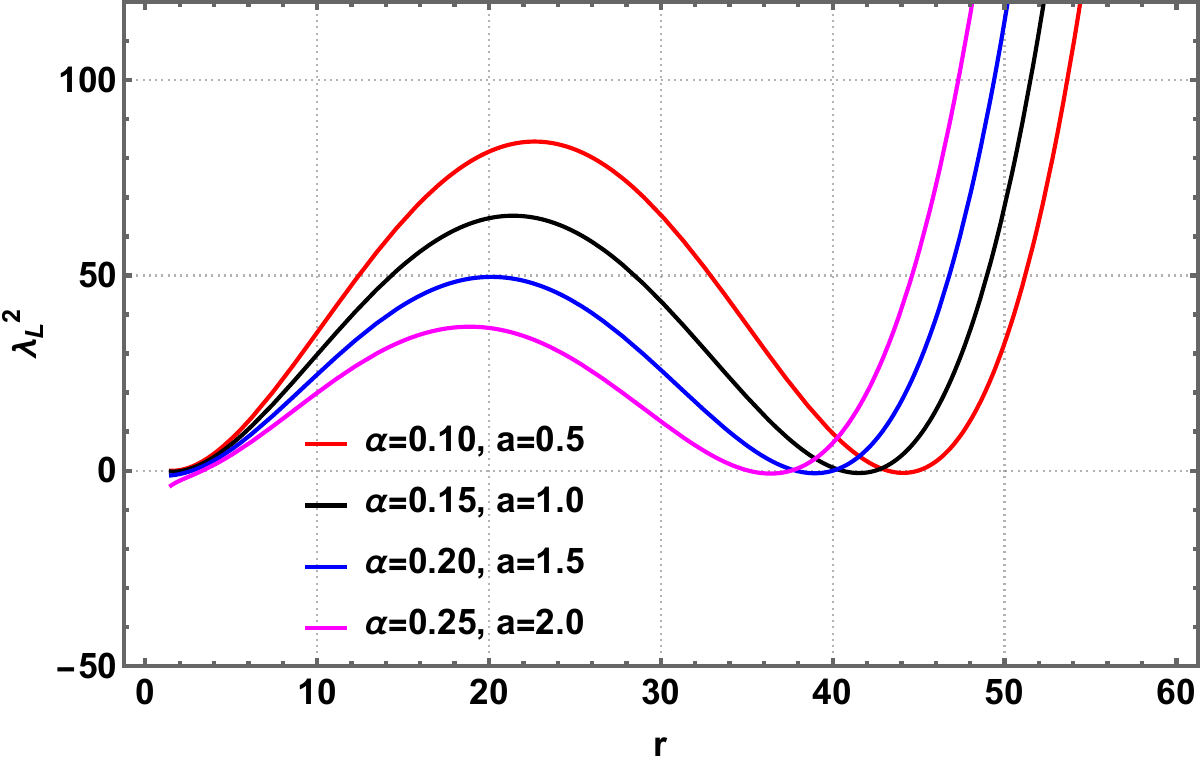}
    \caption{\footnotesize Illustrating the nature of the squared of Lyapunov exponent ($\Lambda^2_L$) as a function of $r$ for different values of $\alpha$ and $a$.}
    \label{fig:lyapunov}
\end{figure}

The above expression (\ref{bb20}) does not explicitly yet show whether the circular null orbits are stable, marginally stable, or unstable. To determine the nature of null orbits, one must assign suitable values to the relevant parameters and analyze the resulting behavior. For that, we have generated Figure \ref{fig:lyapunov} showing the nature of this Lyapunov exponent.

\section{Dynamics of Neutral Test Particles around BH }\label{S4}

In this section, we study the dynamics of neutral test massive particles around the BH solution (\ref{aa4}) and analyze innermost stable circular orbits (ISCO).

Analogue to the previous studies, the Lagrangian density function using the metric (\ref{aa4}) for neutral test particles is given by
\begin{equation}
    \mathcal{L}=\frac{1}{2}\,\left[-f(r)\,\left(\frac{dt}{d\lambda}\right)^2+\frac{1}{f(r)}\,\left(\frac{dr}{d\lambda}\right)^2+r^2\,\left(\frac{d\phi}{d\lambda}\right)^2\right].\label{cc1}
\end{equation}

Here $2\,\mathcal{L}=-1$ for the massive neutral test particles. Here, also two are two conserved quantities associated with coordinates $(t, \phi)$ and are given in Eqs.~(\ref{bb1a})--(\ref{bb2}). Thereby, eliminating $dt/d\lambda$ and $d\phi/d\lambda$ and after simplification results the following one-dimensional equation of motion for the $r$ coordinate as,
\begin{equation}
    \left(\frac{dr}{d\lambda}\right)^2+U_\text{eff}(r)=\mathrm{E}^2.\label{cc2}
\end{equation}
Here $U_\text{eff}(r)$ is the effective potential governing the dynamics of massive neutral test particles and is given by
\begin{equation}
    U_\text{eff}(r)=\left(1+\frac{\mathrm{L}^2}{r^2}\right)\,f(r)=\left(1+\frac{\mathrm{L}^2}{r^2}\right)\,\left[1-\alpha-\frac{2\,M}{r}+\frac{(Q^2-a/2)}{r^2}-\frac{\mathrm{N}}{r^{3\,\omega+1}}-\frac{\Lambda}{3}\,r^2\right].\label{cc3}
\end{equation}

For innermost stable circular orbits (ISCO), the following conditions must be satisfied:
\begin{equation}
    \mathrm{E}^2=U_\text{eff}(r),\quad\quad U'_\text{eff}(r)=0,\quad\quad U''_\text{eff}(r) \geq 0,\label{cc4}
\end{equation}
where prime denotes partial derivative w. r. to $r$.

Thereby, using the effective potential given in Eq.~(\ref{cc3}) into the condition $U'_\text{eff}(r)=0$ and after simplification, we find the specific angular momentum $\mathrm{L}_\text{specific}$ of neutral test particles as,
\begin{equation}
\mathrm{L}_\text{specific}=r\,\sqrt{\frac{\frac{M}{r}- \frac{(Q^2 -a/2)}{r^2} +\dfrac{(3\,\omega + 1)\,\mathrm{N}}{2\,r^{3\,\omega +1}} - \dfrac{\Lambda}{3}\,r^2}{1 - \alpha - \dfrac{3\,M}{r} + \dfrac{(2\,Q^2-a)}{r^2}-\dfrac{(3\,\omega + 3)\,\mathrm{N}}{2\,r^{3\,\omega + 1}}}}.\label{cc5}
\end{equation}
While using the condition $\mathrm{E}^2=U_\text{eff}(r)$, we find the specific energy $\mathrm{E}_\text{specific}$ given by
\begin{equation}
    \mathrm{E}_\text{specific}=\pm\,\frac{\left(1-\alpha-\frac{2\,M}{r}+\frac{(Q^2-a/2)}{r^2}-\frac{\mathrm{N}}{r^{3\,\omega+1}}-\frac{\Lambda}{3}\,r^2\right)}{\sqrt{1 - \alpha - \dfrac{3\,M}{r} + \dfrac{(2\,Q^2 -a)}{r^2}-\dfrac{(3\,\omega + 3)\,\mathrm{N}}{2\,r^{3\,\omega + 1}}}}.\label{cc6}
\end{equation}

These physical quantities (\ref{cc5})--(\ref{cc6}) are influenced by various geometric and physical parameters. These include the string cloud parameter $\alpha$, the quantum deformation parameter $a$, the electric charge $Q$, the normalization constant $\mathrm{N}$ and state parameter $w$ of the quintessence-like fluid. Additionally, the BH mass $M$ and the cosmological constant $\Lambda$ modifies these variables. 

For a particular state parameter, $w=-2/3$, the specific angular momentum and energy of the neutral test particles reduces as,
\begin{equation}
    \mathrm{L}_\text{specific}=r\,\sqrt{\frac{\frac{M}{r}- \frac{(Q^2 -a/2)}{r^2} -\dfrac{\mathrm{N}}{2}\,r-\dfrac{\Lambda}{3}\,r^2}{1 - \alpha - \dfrac{3\,M}{r} + \dfrac{(2\,Q^2-a)}{r^2}-\dfrac{\mathrm{N}}{2}\,r}}\quad,\quad \mathrm{E}_\text{specific}=\pm\,\frac{\left(1-\alpha-\frac{2\,M}{r}+\frac{(Q^2-a/2)}{r^2}-\mathrm{N}\,r-\frac{\Lambda}{3}\,r^2\right)}{\sqrt{1-\alpha-\dfrac{3\,M}{r} + \dfrac{(2\,Q^2 -a)}{r^2}-\dfrac{\mathrm{N}}{2}\,r}}.\label{cc8}
\end{equation}

Finally, using the third condition, we find the following equation for ISCO radius $r=r_\text{ISCO}$ by setting $w=-2/3$ as ($Q^2_\text{eff}=Q^2-a/2$),
\begin{align}
    &\left((1-\alpha)\,r^2-2\,M\,r+Q^2_\text{eff}-\mathrm{N}\,r^3-\frac{\Lambda}{3}\,r^4\right)\,\left(-4\,M\,r+6\,Q^2_\text{eff}- \frac{2\Lambda}{3}\,r^4\right)+\left(2\,M\,r- 2\,Q^2_\text{eff}-\mathrm{N}\,r^3 - \frac{2\,\Lambda}{3}\,r^4\right)\times\nonumber\\
    &\left((3 - 3\,\alpha)\,r^2 -10\,M\,r+7\,Q^2_\text{eff}- \mathrm{N}\,r^3 + \frac{\Lambda}{3}\,r^4\right)=0.\label{cc7}
\end{align}

\begin{table}[ht!]
\centering
\begin{tabular}{|c|c|cc|cc|}
\hline
\multirow{2}{*}{$\alpha$} & \multirow{2}{*}{$a$} & \multicolumn{2}{c|}{$Q = 0.5$} & \multicolumn{2}{c|}{$Q= 1$} \\
                         &                     & $r_{\mathrm{ISCO}}$ (smallest) & $r_{\mathrm{ISCO}}$ (largest) & $r_{\mathrm{ISCO}}$ (smallest) & $r_{\mathrm{ISCO}}$ (largest) \\
\hline
0.05 & 0.25 & 6.27636 & 121.71 & 5.08725 & 121.73 \\
\hline
0.05 & 0.5  & 6.41776 & 121.706 & 5.34869 & 121.727 \\
\hline
0.05 & 0.75 & 6.5499  & 121.703 & 5.57576 & 121.723 \\
\hline
0.1  & 0.25 & 6.53257 & 114.797 & 5.43659 & 114.82 \\
\hline
0.1  & 0.5  & 6.66667 & 114.793 & 5.67201 & 114.816 \\
\hline
0.1  & 0.75 & 6.79259 & 114.789 & 5.87926 & 114.812 \\
\hline
0.15 & 0.25 & 6.80439 & 107.852 & 5.79435 & 107.878 \\
\hline
0.15 & 0.5  & 6.9317  & 107.847 & 6.00604 & 107.874 \\
\hline
0.15 & 0.75 & 7.05185 & 107.843 & 6.19503 & 107.869 \\
\hline
0.2  & 0.25 & 7.09575 & 100.867 & 6.16288 & 100.897 \\
\hline
0.2  & 0.5  & 7.21688 & 100.862 & 6.35359 & 100.892 \\
\hline
0.2  & 0.75 & 7.33179 & 100.857 & 6.52622 & 100.887 \\
\hline
\end{tabular}
\caption{\footnotesize ISCO radius $r=r_{\mathrm{ISCO}}$ for different of CS parameter $\alpha$, the quantum-corrected deformation parameter $a$, and the electric charge $Q$. Here, we set $M=1, \Lambda=-0.003, w=-2/3, \mathrm{N}=0.02$.}
\label{tab:ISCO-radius}
\end{table}

The ISCO radius (\ref{cc7}) is therefore influenced by the string cloud parameter $\alpha$, the quantum deformation parameter $a$, the electric charge $Q$, the normalization constant $\mathrm{N}$ of the quintessence-like fluid. Additionally, the BH mass $M$ and the cosmological constant $\Lambda$ modifies this radius. 

Finally, we determine the effective radial force experienced by the neutral particles in the given gravitational field. This force can be determine using the following definition:
\begin{equation}
    \mathcal{F}=-\frac{1}{2}\,\frac{dU}{dr},\label{cc9}
\end{equation}
where the effective potential $U$ is given in Eq.~(\ref{cc3}). 

After simplification results:
\begin{align}
    \mathcal{F}= -\left( \frac{M}{r^2} - \frac{Q^2 - \frac{a}{2}}{r^3} + \frac{N (3\omega + 1)}{2 r^{3\omega + 2}} - \frac{\Lambda}{3} r \right) + \mathrm{L}^2\, \left[ \frac{1 - \alpha}{r^3} - \frac{3M}{r^4} + \frac{2\left(Q^2 - \frac{a}{2}\right)}{r^5} - \frac{3(\omega + 1)}{2} \frac{N}{r^{3\omega + 4}} \right].\label{cc10}
\end{align}

From expression (\ref{cc10}), we observe that the effective radial force is influenced by the string cloud parameter $\alpha$, the quantum deformation parameter $a$, the electric charge $Q$, the normalization constant $\mathrm{N}$ and state parameter $w$ of the quintessence-like fluid. Additionally, the BH mass $M$ and the cosmological constant $\Lambda$ modifies this radius. 

For a particular state parameter, $w=-2/3$, the effective radial force reduces as,
\begin{align}
    \mathcal{F}= -\left( \frac{M}{r^2} - \frac{Q^2 - \frac{a}{2}}{r^3} - \frac{N}{2 } - \frac{\Lambda}{3} r \right) + \mathrm{L}^2\,\left[ \frac{1 - \alpha}{r^3} - \frac{3\,M}{r^4} + \frac{(2\,Q^2 - a)}{r^5} - \frac{1}{2}\,\frac{N}{r^2} \right].\label{cc11}
\end{align}

\section{Thermodynamics of BH } \label{S5}

As stated earlier, BH thermodynamics shows profound links between gravity, quantum theory, and statistical mechanics. The pioneering work of Bekenstein established that black hole entropy is proportional to the horizon area \cite{SH1}, while Hawking showed that black holes emit thermal radiation with a temperature related to their surface gravity \cite{SH2}. The thermodynamic behavior of black holes in Anti-de Sitter (AdS) space, including the Hawking-Page phase transition, was elucidated by Hawking and Page \cite{SH3}, revealing deep connections to quantum field theories via the AdS/CFT correspondence. Later studies uncovered Van der Waals-like phase transitions in charged AdS black holes when the cosmological constant was treated as pressure \cite{SH4}. For a comprehensive review of these developments and the quantum aspects of black hole thermodynamics, see \cite{SH5}.

In this section, we examine the thermodynamic properties of the selected BH solution given by the metric (\ref{aa4}) with the metric function given in Eq.~(\ref{aa6}), deriving key quantities such as the Hawking temperature, Gibbs free energy, and specific heat capacity. We also analyze how geometric and physical parameters influence and modify these thermodynamic variables.

The mass of the BH solution is determined by applying the condition $f(r_{+})=0$ \cite{SH3}
\begin{equation}
    M=\frac{r_{+}}{2}\,\left[1-\alpha+\frac{(Q^2-a/2)}{r^2_{+}}-\frac{\mathrm{N}}{r^{3\,w+1}_{+}}-\frac{\Lambda}{3}\,r^2_{+}\right].\label{ss1}
\end{equation}

In the extended phase space, one can treat the cosmological
constant as thermodynamic pressure and its conjugate quantity as
thermodynamic volume. The definitions are as follows \cite{DK}:
\begin{equation}
    P=-\frac{\Lambda}{8\,\pi},\quad\quad V=\frac{\partial M}{\partial P}.\label{ss2}
\end{equation}
The conventional method for calculating the Hawking temperature $T_{H}$ of the BH is expressed as follows:
\begin{equation}
    T_H=\frac{f'(r_{+})}{4\,\pi}.\label{ss3}
\end{equation}

\begin{figure}[ht!]
    \centering
    \includegraphics[width=0.3\linewidth]{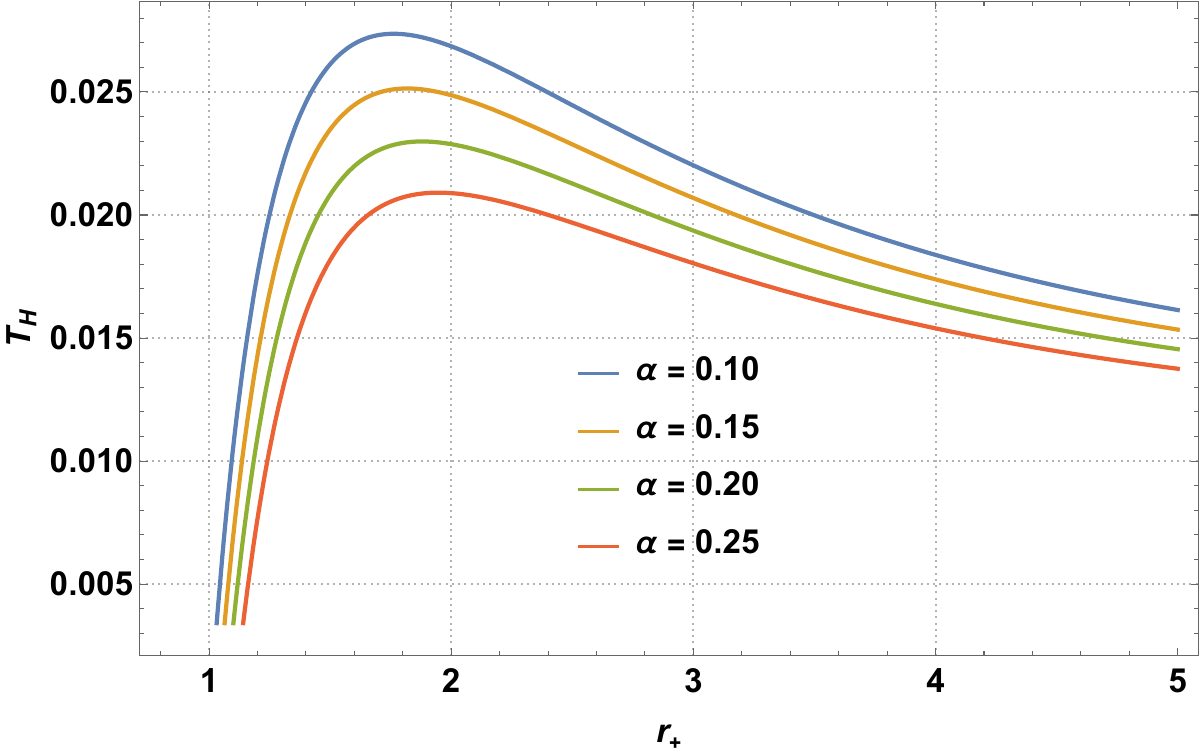}\quad\quad
    \includegraphics[width=0.3\linewidth]{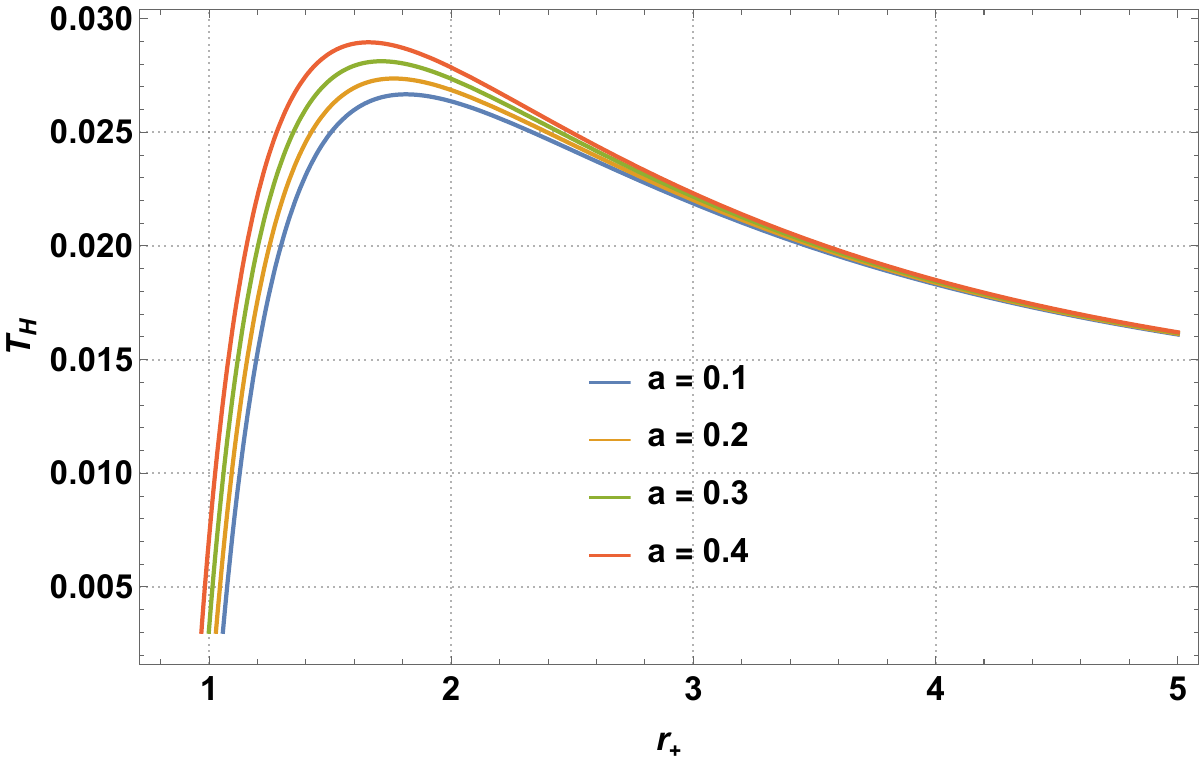}\quad\quad
    \includegraphics[width=0.3\linewidth]{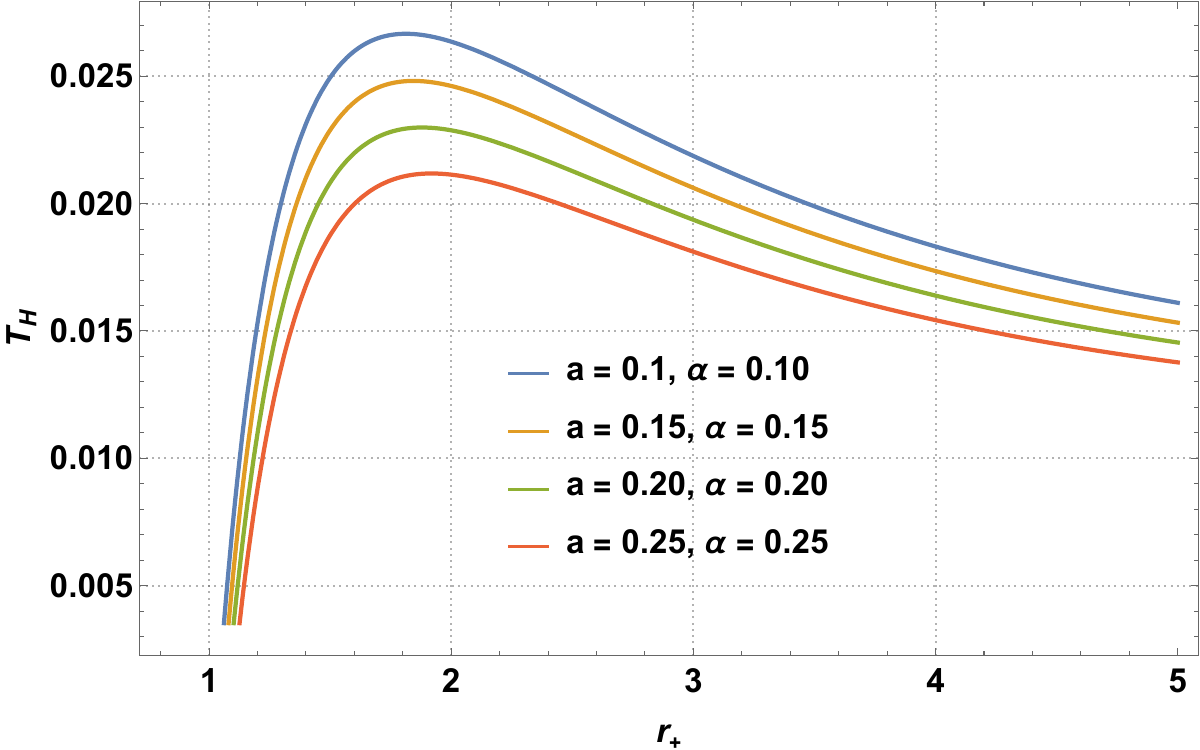}\\
    (a) $a=0.2$  \hspace{5cm} (b) $\alpha=0.1$ \hspace{5cm} (c) 
    \caption{\footnotesize Behavior of the Hawking temperature $T_H$ as a function of horizon radius $r_{+}$ for different values of CoS parameter $\alpha$, the deformation parameter $a$, and their combination. Here, we set $Q=1,\,\mathrm{N}=0.01,\,w=-2/3,\,P=0.01/(8\pi)$.}
    \label{fig:temperature}
\end{figure}

In our case, we find the following expression of the Hawking temperature
\begin{align}
    T_H=\frac{1}{4\,\pi}\,\Bigg[\frac{1 - \alpha}{r_+}
- \frac{Q^2 - \frac{a}{2}}{r_+^3}
+ 3\omega\, \mathrm{N}\,r_+^{-3\omega - 2}
+ 8\pi P\, r_+ \Bigg].\label{ss4}
\end{align}

From expression (\ref{ss4}), we observe that the Hawking temperature is influenced by CoS parameter $\alpha$, the quantum deformation parameter $a$, the electric charge $Q$, the normalization constant $\mathrm{N}$ and state parameter $w$ of the quintessence-like fluid. Additionally, the thermodynamic pressure modifies this temperature. 

The Hawking-Bekenstein entropy is given by \cite{MS}
\begin{equation}
S = \pi\,r_+^2.\label{ss5}
\end{equation}
Considering BH mass $M$ as enthalpy, the Gibbs free energy is obtained as,
\begin{equation}
G=M-T_H\,S= \frac{r_+}{4}\,\left[1 - \alpha+3\, \frac{Q^2 - \frac{a}{2}}{r^2_+}-(2 + 3\omega)\, \mathrm{N} \, r_+^{-3\omega-1}- \frac{8\,\pi}{3}\,P \, r_+^{2}\right].\label{ss6}
\end{equation}

\begin{figure}[ht!]
    \centering
    \includegraphics[width=0.3\linewidth]{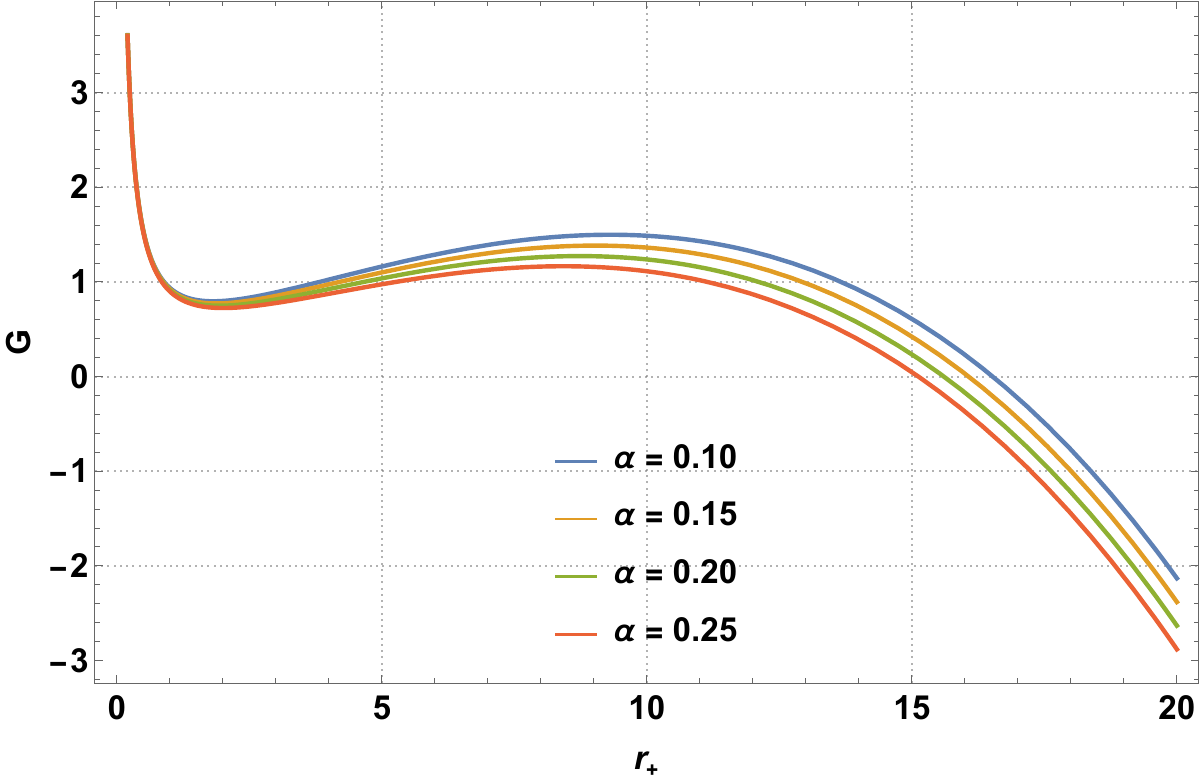}\quad\quad
    \includegraphics[width=0.3\linewidth]{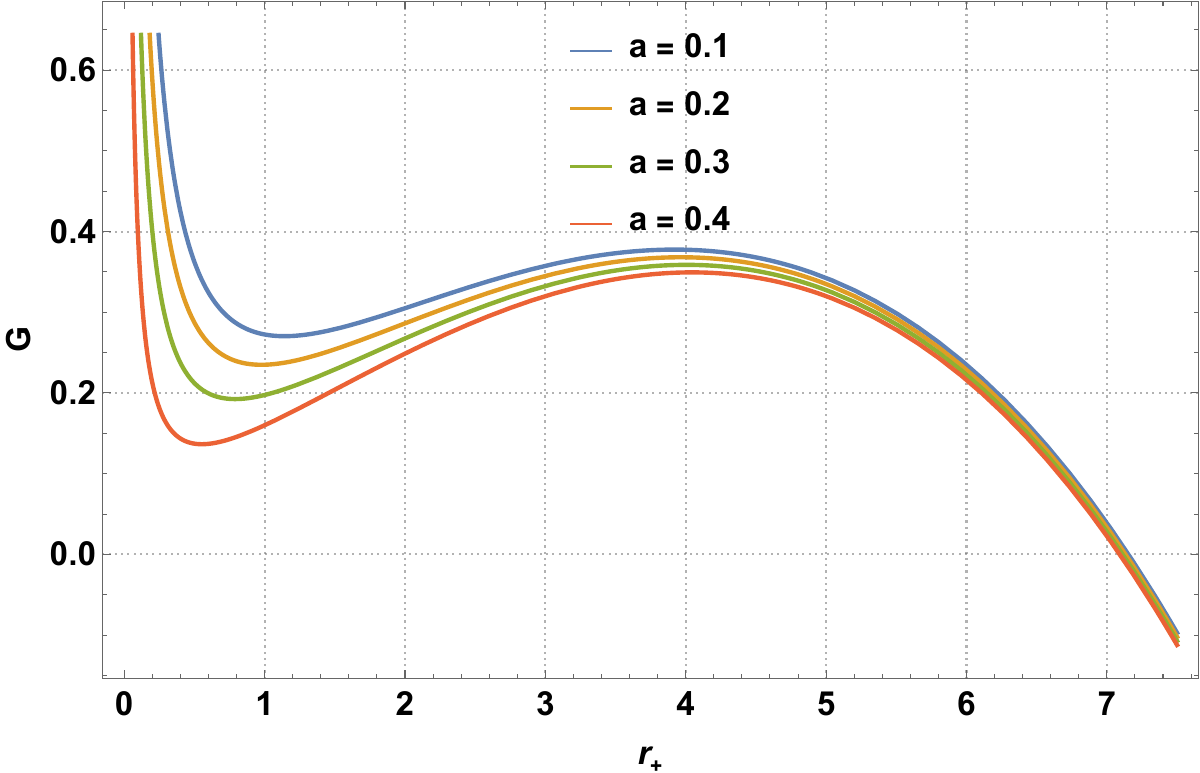}\quad\quad
    \includegraphics[width=0.3\linewidth]{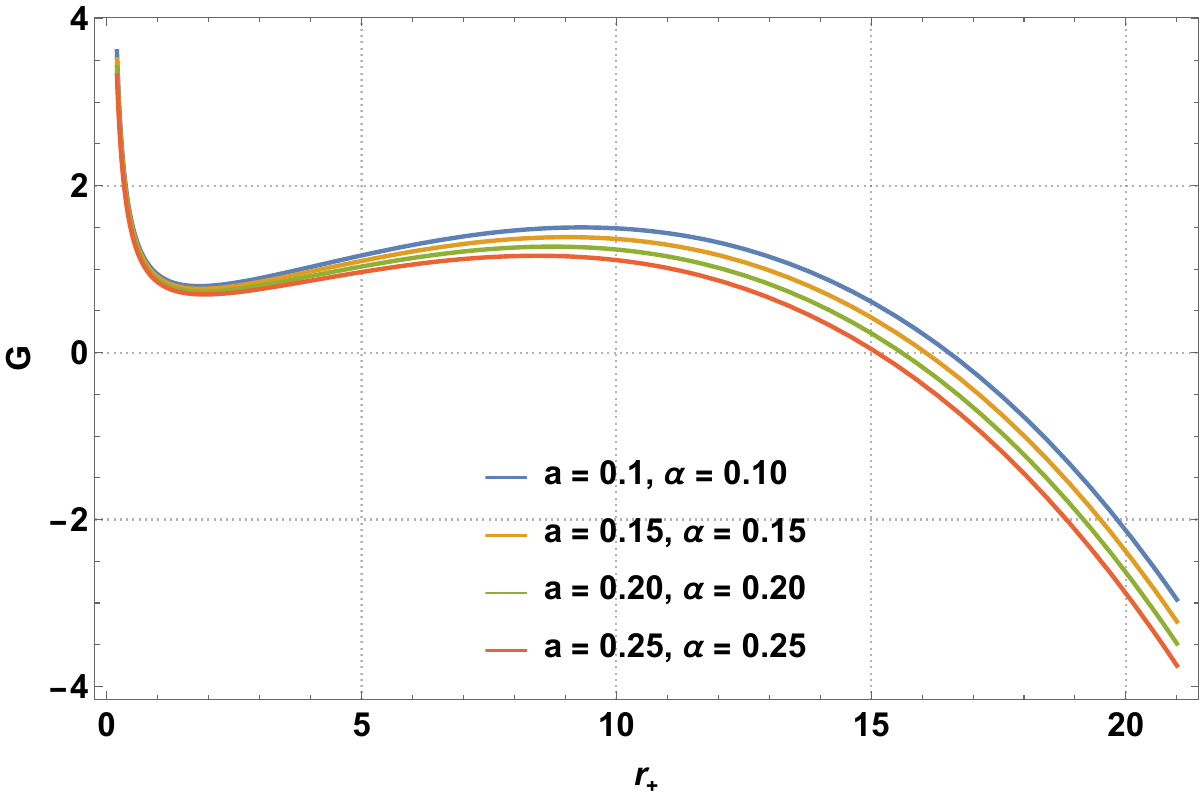}\\
    (a) $a=0.2$  \hspace{5cm} (b) $\alpha=0.1$ \hspace{5cm} (c) 
    \caption{\footnotesize Behavior of the Gibb's free energy $G$ as a function of horizon radius $r_{+}$ for different values of CoS parameter $\alpha$, the deformation parameter $a$, and their combination. Here, we set $Q=1,\,\mathrm{N}=0.01,\,w=-2/3,\,P=0.01/(8\pi)$.}
    \label{fig:Gibbs}
\end{figure}

From expression (\ref{ss6}), we observe that the Gibbs free energy is influenced by CoS parameter $\alpha$, the quantum deformation parameter $a$, the electric charge $Q$, the normalization constant $\mathrm{N}$ and state parameter $w$ of the quintessence-like fluid. Additionally, the thermodynamic pressure modifies this energy.

To find the internal energy of the system, we re-write the BH mass $M$ in terms of the thermodynamic pressure as,
\begin{equation}
    M=\frac{r_{+}}{2}\,\left[1-\alpha+\frac{(Q^2-a/2)}{r^2_{+}}-\frac{\mathrm{N}}{r^{3\,w+1}_{+}}+\frac{8\pi}{3}\,P\,r^2_{+}\right].\label{ss10}
\end{equation}
The thermodynamic volume using Eq. (\ref{ss2}) is obtained as,
\begin{equation}
    V=\frac{4\pi}{3}\,r^3_{+}.\label{ss11}
\end{equation}
Therefore, the internal of the BH system is given by
\begin{equation}
    U=M-P\,V.\label{ss12}
\end{equation}
Substituting $M, V$ using Eqs.~(\ref{ss10}) and (\ref{ss11}), we find the following expression of the internal energy
\begin{equation}
    U=\frac{r_+}{2} \left[
1 - \alpha + \frac{Q^2 - \dfrac{a}{2}}{r_+^2} - \frac{\mathrm{N}}{r_+^{3\omega + 1}}
\right].\label{ss13}
\end{equation}

From expression (\ref{ss13}), we observe that the internal energy is influenced by CoS parameter $\alpha$, the quantum deformation parameter $a$, the electric charge $Q$, the normalization constant $\mathrm{N}$ and state parameter $w$ of the quintessence-like fluid.

We now turn our attention to the thermodynamic stability-both local and global of the BH solution. It is well established that the sign of the heat capacity provides a criterion for local stability: a positive heat capacity indicates that the BH is locally thermodynamically stable, while a negative heat capacity signals local instability. On the other hand, the Gibbs free energy governs the global thermodynamic behavior, determining the phase structure and global stability of the system. To proceed, we derive the heat capacity of the BH using the following relation:
\begin{equation}
C= T_H\,\left( \frac{\partial S}{\partial T_H} \right) = T_H\,\frac{\left( \frac{\partial S}{\partial r_{+}} \right)}{\left( \frac{\partial T_H}{\partial r_{+}} \right)}.\label{ss7}
\end{equation}

Using Eqs.~(\ref{ss4}) and (\ref{ss5}), we find the following expression of the specific heat capacity:
\begin{align}
C=2\,\pi\,r^2_{+}\,\frac{\left(1 - \alpha
- \frac{(Q^2 -a/2)}{r_+^2}
+ 3\omega\, \mathrm{N}\,r_+^{-3\omega - 1}
+ 8\pi P\, r^2_+ \right)}{\left(
-1+\alpha
+ \frac{3\left(Q^2 - \frac{a}{2}\right)}{r_+^2}
- 3\,\omega\,(3\,\omega + 2) \mathrm{N} \, r_+^{-3\omega - 1}
+ 8\,\pi\, P\,r^2_{+}
\right)}.\label{ss8}
\end{align}

\begin{figure}[ht!]
    \centering
    \includegraphics[width=0.3\linewidth]{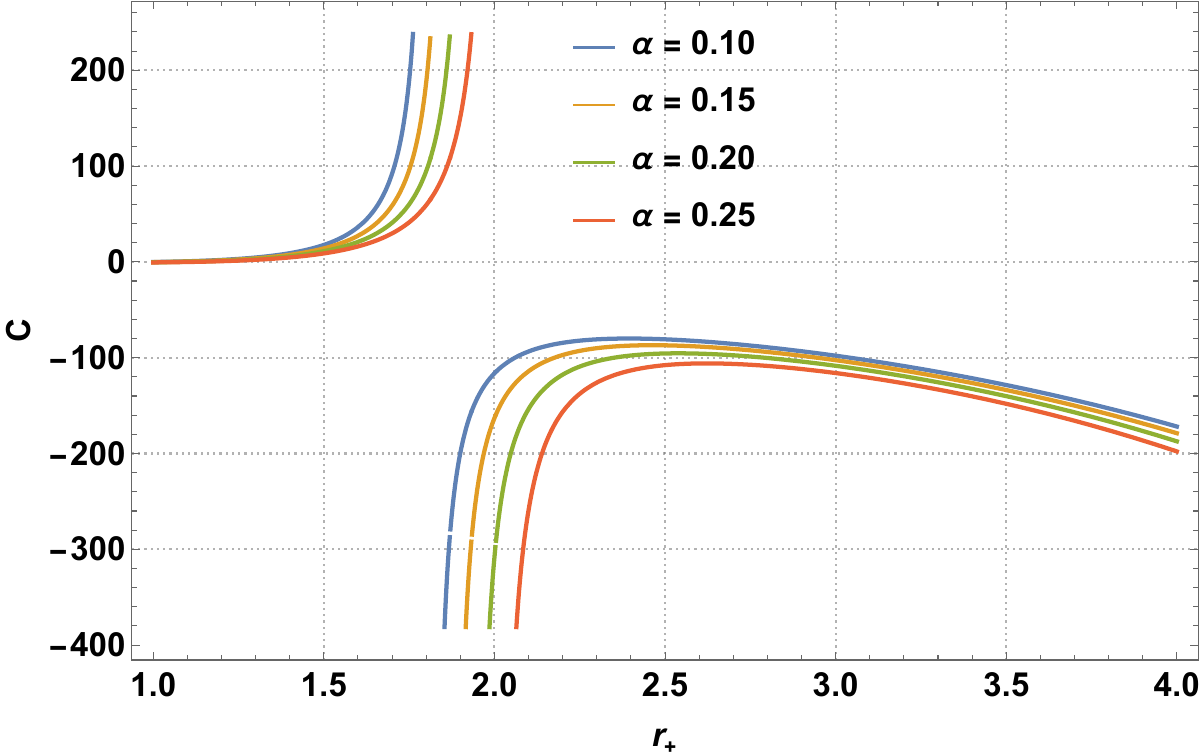}\quad\quad
    \includegraphics[width=0.3\linewidth]{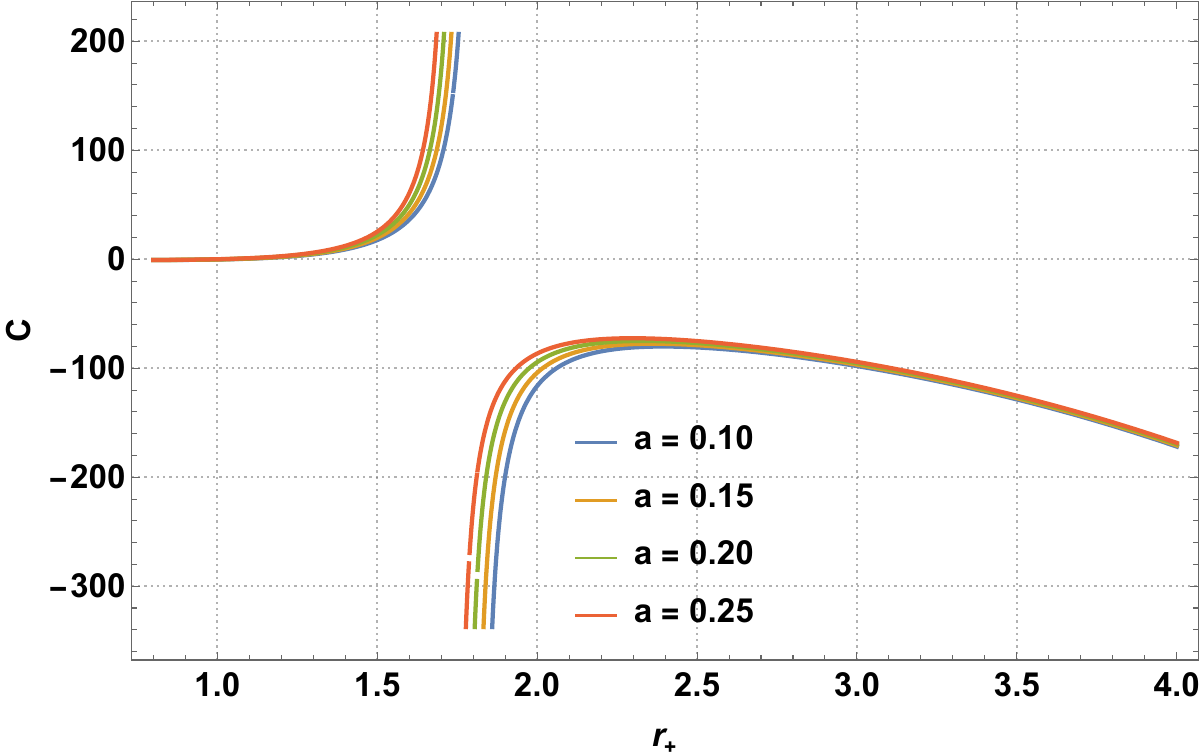}\quad\quad
    \includegraphics[width=0.3\linewidth]{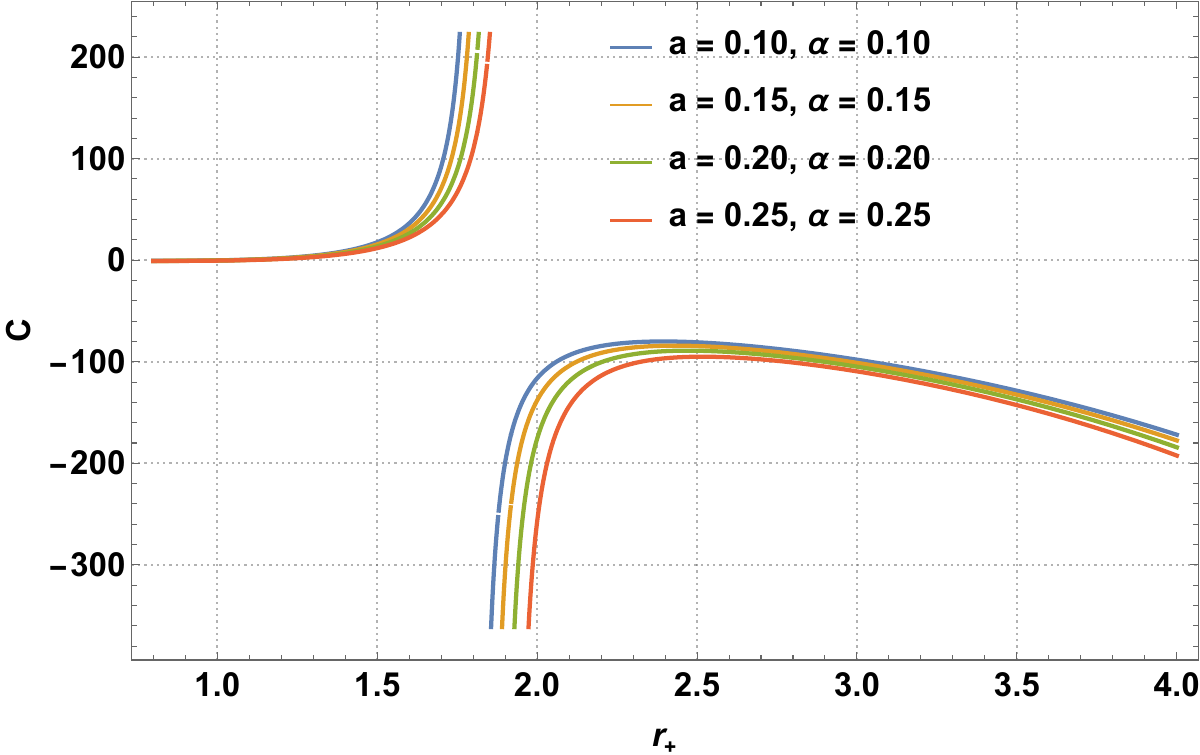}\\
    (a) $a=0.1$  \hspace{5cm} (b) $\alpha=0.1$ \hspace{5cm} (c) 
    \caption{\footnotesize Behavior of the Hawking temperature $T_H$ as a function of horizon radius $r_{+}$ for different values of CoS parameter $\alpha$, the deformation parameter $a$, and their combination. Here, we set $Q=1,\,\mathrm{N}=0.01,\,w=-2/3,\,P=0.01/(8\pi)$.}
    \label{fig:heat}
\end{figure}

From expression (\ref{ss8}), we observe that the specific heat capacity is influenced by CoS parameter $\alpha$, the quantum deformation parameter $a$, the electric charge $Q$, the normalization constant $\mathrm{N}$ and state parameter $w$ of the quintessence-like fluid. Additionally, the thermodynamic pressure modifies it.

For a particular state parameter, $w=-2/3$, the specific heat reduces as,
\begin{align}
C_V =2\,\pi\,r^2_{+}\,\frac{\left(1 - \alpha
- \frac{(Q^2 -a/2)}{r_+^2}
-\mathrm{N}\,r_+
+ 8\pi P\, r^2_+ \right)}{\left(
-1+\alpha
+ \frac{3\left(Q^2 - \frac{a}{2}\right)}{r_+^2}
+ 8\,\pi\, P\,r^2_{+}
\right)}.\label{ss9}
\end{align}

\section{Summary and Conclusions} \label{S6}

In this study, we investigated a novel quantum-corrected deformation of the Reissner–Nordström–Anti-de Sitter black hole solution, coupled with a cloud of strings and surrounded by a quintessence-like fluid. By incorporating a square-root correction term inspired by quantum gravitational effects, the resulting metric captured deviations from classical behavior, particularly in the near-horizon regime. This generalized space-time provided a consistent framework for examining optical properties, test particle dynamics, and thermodynamic behavior. Furthermore, in the absence of quntum correction, the solution smoothly reduced to the classical RN-AdS BH configuration with contributions from both the string cloud and quintessence, thereby confirming the consistency and robustness of the deformation. Overall, this framework laid the foundation for more detailed analyses of observable signatures and potential extensions in the context of quantum-corrected black hole physics.

We explored the optical properties of the BH solution modified by a cloud of strings and a surrounding quintessence-like fluid, demonstrating how geometric and physical parameters shaped the propagation of light. Starting from null geodesics in the equatorial plane, we derived the effective potential governing photon motion and showed its sensitivity to parameters such as the string cloud parameter $\alpha$, electric charge $Q$, quantum correction parameter $a$, normalization constant $\mathrm{N}$, equation-of-state parameter $w$, cosmological constant $\Lambda$, and BH mass $M$. By transforming to the inverse radial coordinate $u = 1/r$, we obtained a second-order nonlinear differential equation describing photon trajectories, highlighting how each parameter perturbed the geometry. We found that increasing $\alpha$ or $a$ enlarged the photon sphere radius $r_{\text{ph}}$ and BH shadow radius $R_s$, as confirmed through both analytical expressions and numerical analysis. The stability of circular null orbits was analyzed using the Lyapunov exponent, which remained positive ($\lambda_L^2 > 0$) for $w = -2/3$, confirming the instability of these orbits. These findings illustrated that quantum corrections and string cloud effects significantly influenced observable quantities such as lensing and shadow size.

We investigated the motion of neutral test particles in the gravitational field of the chosen BH surrounded by a cloud of strings and a quintessence-like fluid. Using the background metric, we formulated the Lagrangian and derived the effective potential $U_{\text{eff}}(r)$, which was significantly influenced by the BH mass $M$, electric charge $Q$, quantum correction parameter $a$, string cloud parameter $\alpha$, normalization constant $N$, equation-of-state parameter $\omega$, and cosmological constant $\Lambda$. These parameters affected the specific energy and angular momentum, thereby altering the conditions for circular motion and orbital stability. By analyzing the extrema of $U_{\text{eff}}(r)$, we obtained expressions for the innermost stable circular orbit (ISCO), showing that the ISCO radius varied with $\alpha$, $a$, and $Q$. An increase in $\alpha$ or $a$ led to a systematic shift in the ISCO radius, while larger $Q$ values resulted in tighter particle confinement. For $\omega = -2/3$, the effective potential and radial force simplified, offering clearer insights into the competing gravitational, electric, and cosmological influences.

The thermodynamic analysis of the considered BH solution shown complex interplay between geometric deformations, matter field contributions, and extended phase space effects, resulting in diverse stability and phase transition behaviors. The Hawking temperature profile showed that the string cloud parameter $\alpha$ and quantum deformation parameter $a$ acted as key regulators-where increasing $\alpha$ suppressed $T_H$ at small horizon radii, while $a$ either increased or decreased $T_H$ depending on its relation to the electric charge $Q$. The quintessence-like fluid, characterized by parameters $\mathrm{N}$ and $w$, introduced non-monotonic temperature behavior, suggesting possible BH remnants for specific parameter ranges. In the Gibbs free energy profile, these parameters shifted global minima and altered slopes, affecting the Hawking–Page transition and stability criteria. Internal energy analysis confirmed that pressure $P$ influenced enthalpy but not internal energy directly, highlighting matter contributions to the energy content. The specific heat capacity $C$ distinguished stable ($C>0$) from unstable ($C<0$) phases, with divergence points marking second-order phase transitions. For $w = -2/3$, the heat capacity simplified, revealing that larger $\mathrm{N}$ values shifted transition points toward larger horizon radii. Overall, the thermodynamic behavior was strongly shaped by $\alpha$, $a$, $Q$, and $\mathrm{N}$, enabling control over phase structure and thermal stability. In the limits $a \to 0$, $\alpha \to 0$ and $\mathrm{N} \to 0$, the model reduced smoothly to the standard charged AdS BH, confirming that the observed complexity arose from the additional corrections. Stability was maintained only within specific parameter ranges, beyond which the system transitioned to unstable or non-physical states.

This study offers a foundation for exploring observational and theoretical implications of quantum-corrected black holes. Future work may include detailed analysis of perturbations, and quasinormal modes to constrain quantum corrections using observational data. Extending the model to include spinning or charged test particles, or exploring higher-order corrections in the thermodynamic phase space, could reveal richer dynamical and thermodynamic behavior. Investigations into holographic heat engines or Joule-Thomson expansions \cite{GM1,GM2,GM3,GM4,GM5} using this background may also provide new insights into BH thermodynamics from the AdS/CFT viewpoint. Embedding the framework into specific quantum gravity models may also provide deeper insights into the microscopic origin of BH entropy and stability.

{\footnotesize
\section*{Acknowledgments}

F.A. acknowledges the Inter University Centre for Astronomy and Astrophysics (IUCAA), Pune, India for granting visiting associateship.

\section*{Data Availability}

No data were generated or created in this article.

\section*{Conflicts of interest statement}

The authors declare(s) no conflicts of interest.

\section*{Funding}

This research has not received any funding.

}

\end{document}